\documentclass[journal, onecolumn, 12pt, draftclsnofoot]{IEEEtran}

\usepackage{amsfonts}
\usepackage{amsmath}
\usepackage{booktabs}
\usepackage{multirow}
\usepackage{array}
\usepackage{bbm}   
\usepackage{color}  
\usepackage{amssymb} 
\usepackage{algorithm}
\usepackage{algpseudocode}
\usepackage{graphicx}  
\usepackage{cite}   
\usepackage{subfigure}

\begin{document}

\bibliographystyle{IEEEtran}

\title{Learn to Compress CSI and Allocate Resources in Vehicular Networks}
%
%
%

\author{Liang~Wang,~\IEEEmembership{Member,~IEEE,}
        Hao~Ye,~\IEEEmembership{Student Member,~IEEE,}
        Le~Liang,~\IEEEmembership{Member,~IEEE,}
        and~Geoffrey~Ye~Li,~\IEEEmembership{Fellow,~IEEE}
\thanks{Liang Wang is with the Key Laboratory of Modern Teaching Technology, Ministry of Education, Xi'an 710062, China, and also with the School of Computer Science, Shaanxi Normal University, Xi'an 710119, China (e-mail: wangliang@snnu.edu.cn).}

\thanks{Hao Ye, Le Liang and Geoffrey Ye Li are with the School
of Electrical and Computer Engineering, Georgia Institute of Technology, Atlanta,
GA, 30332 USA (e-mail: \{yehao, lliang\}@gatech.edu; liye@ece.gatech.edu).}
}

%

\maketitle

\begin{abstract}
Resource allocation has a direct and profound impact on the performance of vehicle-to-everything (V2X) networks. 
In this paper, we develop a hybrid architecture consisting of centralized decision making and distributed resource sharing (the C-Decision scheme) to maximize the long-term sum rate of all vehicles. To reduce the network signaling overhead, each vehicle uses a deep neural network to compress its observed information that is thereafter fed back to the centralized decision making unit. The centralized decision unit employs a deep Q-network to allocate resources and then sends the decision results to all vehicles. We further adopt a quantization layer for each vehicle that learns to quantize the continuous feedback. In addition, we devise a mechanism to balance the transmission of vehicle-to-vehicle (V2V) links and vehicle-to-infrastructure (V2I) links. To further facilitate distributed spectrum sharing, we also propose a distributed decision making and spectrum sharing architecture (the D-Decision scheme) for each V2V link. Through extensive simulation results, we demonstrate that the proposed C-Decision and D-Decision schemes can both achieve near-optimal performance
and are robust to feedback interval variations, input noise, and feedback noise.
\end{abstract}


%
\IEEEpeerreviewmaketitle

\section{Introduction}

\IEEEPARstart{C}{onnecting} vehicles on the road as a dynamic communication network, commonly known as vehicle-to-everything (V2X) networks, is gradually becoming a reality to make our daily experience on wheels safer and more convenient \cite{16V2XService}. V2X enabled coordination among vehicles, pedestrians, and other entities on the road can alleviate traffic congestion, improve road safety, in addition to providing ubiquitous infotainment services \cite{17V2XStandard,17V2PHY,19V2Net}. Recently, the 3rd generation partnership project (3GPP) begins to support V2X services in the long-term evolution (LTE) \cite{SimScenario} and further the fifth
generation (5G) mobile communication networks \cite{V2X5G}. Cross-industry alliance has also been founded, such as the 5G automotive association (5GAA), to push development, testing, and deployment of V2X technologies.

Due to high mobility of vehicles and complicated time-varying communication environments, it is very challenging to guarantee the diverse quality-of-service (QoS) requirements in vehicular networks, such as extremely large capacity, ultra reliability, and low latency \cite{19LowDelayEC}. To address such issues, efficient resource allocation for spectrum sharing becomes necessary in the V2X scenario.
Existing works on spectrum sharing in vehicular networks can be mainly categorized into two classes: centralized schemes \cite{17D2DRA,18graphRA,19NOMAV2X,19segMACV2X} and distributed approaches \cite{11DistCHAss,16DistV2V}. For the centralized schemes, decisions are usually made centrally at a given node, such as the head in a cluster or the base station (BS) in a given coverage area. Novel graph-based resource allocation schemes have been proposed in \cite{17D2DRA} and \cite{18graphRA} to maximize the vehicle-to-infrastructure (V2I) capacity, exploiting the slow fading statistics of channel state information (CSI). In \cite{19NOMAV2X}, an interference hyper-graph based resource allocation scheme has been developed in the non-orthogonal multiple access (NOMA)-integrated V2X scenario with the distance, channel gain, and interference known in each vehicle-to-vehicle (V2V) and V2I group. In \cite{19segMACV2X}, a segmentation medium access control (MAC) protocol has been proposed in large-scale V2X networks, where the location information of vehicles is updated. In these schemes, the decision making node needs to acquire accurate CSI, interference information of all the V2V links, and each V2V link's transmit power to make spectrum sharing decisions. However, reporting all such information from each V2V link to the decision making node poses a heavy burden on the feedback links, and even becomes infeasible in practice.

As for distributed schemes \cite{11DistCHAss,16DistV2V}, each V2V link makes its own decision with partial or little knowledge of other V2V links.
In \cite{11DistCHAss}, a distributed shuffling based Hopcroft-Karp algorithm has been devised to handle the subchannel allocation in V2V communications with one-bit CSI broadcasting. In \cite{16DistV2V}, the spatio-temporal traffic pattern has been exploited for distributed load-aware resource allocation for V2V communications with slowly varying channel information. In these methods, V2V links may exchange partial or none channel information with their neighbors before making a decision. However, each V2V link can only observe partial information of its surrounding environment since it is geographically apart from other V2V links in the V2X scenario. This may leave some channels overly congested while others underutilized, leading to substantial performance degradation.

Notably, the above works usually rely on some levels of channel information, such as channel gain, interference, locations and so on. This kind of channel information is usually hard to obtain perfectly in practical wireless communication systems, which is even challenging in the V2X scenario. Fortunately, machine learning enables wireless communications systems to learn their surroundings and feed critical information back to the BS for resource allocation. In particular, reinforcement learning (RL) can make decisions to maximize long-term return in the sequential decision problems, which has gained great success in various applications, such as AlphaGo \cite{16AlphaGo}.
Inspired by its remarkable performance, the wireless community is increasingly interested in leveraging machine learning for the physical layer and resource allocation design \cite{17DLPHY,19DLPHYComm,18DLCHEstDet,18E2EoCHModel, 17MLnextG,17Intelli5G,18multiCHaccess,19RLfog,liang2019deep}. In particular, machine learning for future vehicular networks has been discussed in \cite{18MLv2x} and \cite{19IntelliV2V}.
In \cite{19HYeBroadcast}, each V2V link is treated as an agent to ensure the latency constraint is satisfied while minimizing interference to V2I link transmission. In \cite{19MARL}, a multi-agent RL-based spectrum sharing scheme has been proposed to promote the payload delivery rate of V2V links while improving the sum capacity of V2I links. A dynamic RL scheduling algorithm has been developed to solve the network traffic and computation offloading problems in vehicular networks \cite{19RLoffload}.

In order to fully exploit the advantages of both centralized and distributed schemes while alleviating the requirement on CSI for spectrum sharing in vehicular networks, we propose an RL-based resource allocation scheme with learned feedback. In particular, we devise a distributed CSI compression and centralized decision making architecture to maximize the sum rate of all V2V links in the long run. 
In this architecture, each V2V link first observes the state of its surrounding channels and adopts a deep neural network (DNN) to learn what to feed back to the decision making unit, such as the BS, instead of sending all observed information directly. To maximize the long-term sum rate of all links, the BS then adopts deep reinforcement learning to allocate spectrum for all V2V links. To further reduce feedback overhead, we adopt a quantization layer in each vehicle's DNN and learn how to quantize the continuous feedback. Besides, to further facilitate distributed spectrum sharing, we devise a distributed spectrum sharing architecture to let each V2V link make its own decision locally. The contributions of this paper are summarized as follows.
\begin{itemize}
\item We leverage the power of DNN and RL to devise a centralized decision making and distributed implementation architecture for vehicular spectrum sharing that maximizes the long-term sum rate of all vehicles. We use a weighted sum rate reward to balance V2I and V2V performance dynamically. 
  \item We exploit the DNN at each vehicle to compress local observations, which is further augmented by a quantized layer, to reduce network signaling overhead while achieving desirable performance. 
  \item We also develop a distributed decision making architecture that allows spectrum sharing decisions to be made at each vehicle locally and binary feedback is designed for signaling overhead reduction.
  \item Based on extensive computer simulations, we demonstrate both of the proposed architectures can achieve near-optimal performance and are robust to feedback interval variations, input noise, and feedback noise.
  In addition, the optimal number of continuous feedback and feedback bits for each V2V link are presented that strike a balance between signaling overhead and performance loss. 
\end{itemize}

The rest of this paper is organized as follows. 
The system model is presented in Section II. 
Then, the BS aided spectrum sharing architecture, including distributed CSI compression and feedback, centralized resource allocation and quantized feedback, is introduced in Section III. 
The distributed decision making and spectrum sharing architecture is discussed in Section IV. 
Simulation results are presented in Section V. 
Finally, conclusions are drawn in Section VI.

\section{System Model}
We consider a vehicular communication network with $N$ cellular users (CUs) and  ${K}$ pairs of coexisting device-to-device (D2D) users, where all devices are equipped with a single antenna. 
Let $\mathcal{K} = \left\{1,2,...,K \right\}$ and $\mathcal{N} = \left\{1,2,...,N \right\}$ denote the sets of all D2D pairs and CUs, respectively. 
Each pair of D2D users exchange important and short messages, such as safety-related information via establishing a V2V link
while each CU uses a V2I link to support bandwidth-intensive applications, such as social networking and video streaming. In order to ensure the QoS of the CUs, we assume all V2I links are assigned orthogonal radio resources. Without loss of generality, we assume that each CU occupies one channel for its uplink transmission. To improve the spectrum utilization efficiency, all V2V links share the spectrum resource with V2I links. Therefore, $\mathcal{N}$ is also referred to as the channel set.

Denote the channel power gain from the $n$-th CU to the BS on the $n$-th channel, i.e., the $n$-th V2I link, by $g_{n}\left[n\right]$. Let $h_{k,B} \left[n\right]$ represent the cross channel power gain from the transmitter of the $k$-th V2V link to the BS on the $n$-th channel. The received signal-to-interference-plus-noise-ratio (SINR) of the $n$-th V2I link can be expressed as
\begin{equation}
 \gamma^c_n \left[ n \right] = \frac{P^c_n {g_{n}\left[ n \right]}}{\sum_{k=1}^{K}{{\rho_k \left[ n \right]}P^d_k{h_{k,B}\left[n\right]}}  + \sigma^2 },
\label{Eq1}
\end{equation}
where $P^c_n$ and $P^d_k$ refer to the transmit powers of the $n$-th V2I link and the $k$-th D2D pair, respectively, $\sigma^2$ represents the noise power, and ${\rho_k \left[n\right]} \in \left\{0,1 \right\}$ is the channel allocation indicator with ${\rho_k \left[n\right]} = 1 $ if the $k$-th D2D user pair chooses the $n$-th channel and ${\rho_k \left[n\right]} = 0$ otherwise. We assume each D2D pair only occupies one channel, i.e., $\sum_{n=1}^{N} \rho_k\left[n\right] \le 1$. Then, the capacity of the $n$-th V2I link on the $n$-th channel can be written as
\begin{equation}
C^c_n \left[n  \right]= B \log_2\left(1 + \gamma^c_n \left[ n \right] \right),
\label{Eq2}
\end{equation}
where $B$ denotes the channel bandwidth.

Similarly, $h_k[n]$ denotes the channel power gain of the $k$-th V2V link on the $n$-th channel. Meanwhile, $h_{l,k}\left[n\right]$ denotes the cross channel power gain from the transmitter of the $l$-th D2D pair to the receiver of the $k$-th D2D pair on the $n$-th channel. Denote the cross channel power gain from the $n$-th CU to the receiver of the $k$-th D2D pair on the $n$-th channel by $g_{n, k}\left[ n \right]$. Then, the SINR of the $k$-th V2V link over the $n$-th channel can be written as
 \begin{equation}
\gamma^d_k \left[ n \right]= \frac{{\rho_k \left[n\right]}  P_k^d {h_{k} \left[n\right] } }{I_k \left[n\right] + \sigma^2},
\label{Eq3}
\end{equation}
where the interference power for the $k$-th V2V link $I_k \left[n\right]$ is
 \begin{equation}
I_k \left[n\right] = \sum_{l \neq k}^{K}{{\rho_l\left[n\right]}P_l^d h_{l,k} \left[n\right] }  + P_n^c g_{n,k} \left[ n\right].
\label{Eq4}
\end{equation}
In (\ref{Eq4}), the terms $\sum_{l \neq k}^{K}{{\rho_l\left[n\right]}P_l^d h_{l,k} \left[n\right] }$ and $P_n^c g_{n,k} \left[ n\right]$ refer to the interference of the other V2V links and the V2I link on the $n$-th channel, respectively.
Hence the capacity of the $k$-th V2V link on the $n$-th channel can be written as
\begin{equation}
C_k^d \left[n\right] = B \log_2\left(1 + \gamma^d_k \left[ n \right] \right).
\label{Eq5}
\end{equation}

In the V2X networks, a naive distributed approach will allow each V2V link to select a channel independently such that its own data rate is maximized. However, local rate maximization often leads to suboptimal global performance due to the interference among different V2V links. On the other hand, the BS in the V2X scenario has enough computational and storage resources to achieve efficient resource allocation. With the help of machine learning, we propose a centralized decision making scheme based on compressed information learned by each individual V2V link distributively.

In order to achieve this goal, each V2V link first learns to compress local observations, including the channel gain, the observed interference from other V2V links and V2I link, transmit power, etc., and then feeds the compressed information back to the BS. According to feedback information from all V2V links, the BS will make optimal decisions for all V2V links using RL. Then, the BS broadcasts the decision result to all V2V links.


\section{BS Decision based Spectrum Sharing Architecture}
As shown in Fig. \ref{Fig1}, we adopt the deep RL approach for resource allocation. In this section, we first design the DNN architecture of each V2V link and the deep Q-network (DQN) for centralized control at the BS, respectively. Then, we propose the centralized decision making and distributed spectrum sharing architecture, termed C-Decision scheme. Finally, we introduce the binary feedback design for information compression.

\subsection{V2V DNN Design}
\begin{figure}
\begin{center}
\includegraphics[width=0.6\textwidth]{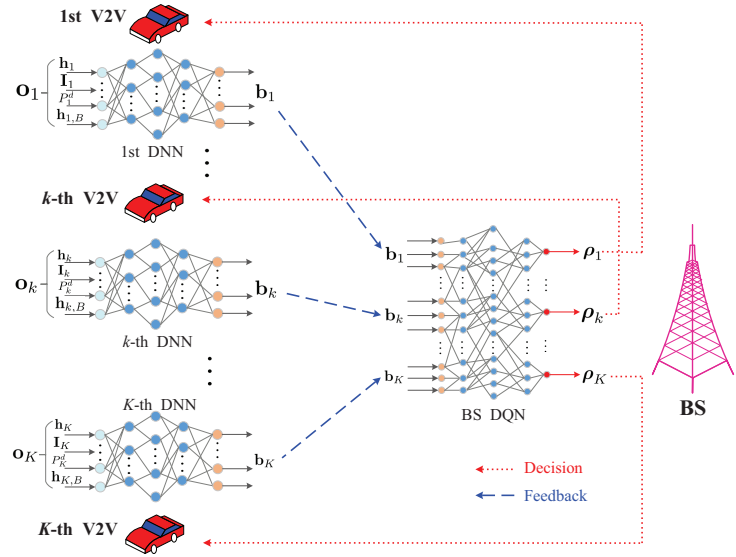}
\caption{Neural network architecture for V2V links and the BS in the C-Decision scheme.}
\label{Fig1}
\end{center}
\end{figure}

Here, we discuss the DNN at each V2V link to compress local observation for feedback. As shown in Fig. \ref{Fig1}, each V2V link $k$ first observes its surroundings, and obtains its transmission power, the current channel gains and interference powers of all channels, which are denoted as $\mathbf{h}_k = \left(h_k\left[ 1 \right], ..., h_k\left[ n \right], ..., h_k\left[ N \right]\right)$ and $\mathbf{I}_k = \left(I_k\left[ 1 \right], ..., I_k\left[ n \right], ..., I_k\left[ N \right] \right)$, respectively. Here, $I_k\left[ n \right]$ refers to the aggregated interference powers at the $k$-th V2V link on the $n$-th channel as shown in (\ref{Eq4}). To consider the impact of V2V links on V2I links, the observation of the $k$-th V2V also needs to include the cross channel gain from the $k$-th V2V link to all V2I links, such as $h_{k,B} \left[ n \right], \forall n \in \mathcal{N}$. Then, the observation of the $k$-th V2V can be written as
\begin{equation}
\mathbf{o}_k = \left\{ \mathbf{h}_k, \mathbf{I}_k, P^d_k, \mathbf{h}_{k,B} \right\},
\label{Eq6}
\end{equation}
where $\mathbf{h}_{k,B} = \left(h_{k, B} \left[1\right], ..., h_{k, B}\left[n\right], ..., h_{k, B}\left[N\right] \right)$. Here, the channel information $\mathbf{h}_k$ can be accurately estimated by the receiver of the $k$-th V2V link and we assume it is also available at the transmitter through delay-free feedback \cite{nasir2018deep}. Similarly, the received interference power over all channels $\mathbf{I}_k$ can be measured at the $k$-th V2V receiver. Each V2V transmitter knows its transmit power $P^d_k$. Besides, the vector $\mathbf{h}_{k,B}$ can be estimated at the BS and then broadcast to all V2V links in its coverage, which incurs a small signaling overhead \cite{19MARL}.


Then, the local observation, $\mathbf{o}_k$, is compressed using the DNN at each V2V link. The compressed information, $\mathbf{b}_k$, which is the output of the DNN, is fed back to the DQN at the BS. To limit overhead on information feedback, each V2V link only reports the compressed information vector, $\mathbf{b}_k$, instead of $\mathbf{o}_k$ to the BS.
Here, $\mathbf{b}_k = \left\{ b_{k,j} \right\}$ is also known as the feedback vector of the $k$-th V2V link and the term $b_{k,j}, \forall j \in \left\{1,2,...,{N}_k\right\}$ refers to the $j$-th feedback element of the $k$-th V2V, where ${N}_k$ denotes the number of feedback learned by the $k$-th V2V link. All V2V links aim at maximizing their global sum rate in the long run while minimizing the feedback information $\mathbf{b}_k$. Therefore, the parameters of the DNNs at all V2V links and those of the DQN will be jointly determined to maximize the sum rate of the whole V2X network.
\subsection{Deep Q-Network at the BS}
To make a proper resource sharing decision, we introduce the deep RL architecture at the BS as shown in Fig. \ref{Fig1}. In order to maximize the long-term sum rate of all links, we resort to the RL technique by treating the BS as the agent. In the RL, an agent interacts with its surroundings, named as the environment, via taking actions, and then observes a corresponding numerical reward from the environment. The agent's goal is to find optimal actions so that the expected sum of rewards is maximized. Mathematically, the RL can be modelled as a Markov decision process (MDP). At each discrete time slot $t$, the agent observes the current state $S_t$ of the environment from the state space $\mathcal{S}$ and then chooses an action $A_t$ from the action space $\mathcal{A}$ and one time step later obtains a reward $R_{t+1}$.
Then, the environment evolves to the next state $S_{t+1}$, with the transition probability $p\left(s',r|s, a\right) \triangleq \Pr \left\{S_{t+1} = s', R_{t+1} = r | S_t = s, A_t = a \right\}$.

The BS treats all the learned feedback as the current state $\mathbf{s}$ of the agent's environment, which can be expressed as:
\begin{equation}
\mathbf{s} = \left\{\mathbf{b}_1,\mathbf{b}_2,...,\mathbf{b}_K \right\}, \forall k \in \mathcal{K}.
\label{Eq7}
\end{equation}
Then, the action of the BS is to determine the values of the channel indicators, $\rho_k \left[n \right]$, for each V2V link. Thus, we define the action $\mathbf{a}$ of the BS as
\begin{equation}
\mathbf{a} = \left\{\boldsymbol{\rho}_1, ... , \boldsymbol{\rho}_k,  ..., \boldsymbol{\rho}_K \right\} ,\forall k \in \mathcal{K},
\label{Eq8}
\end{equation}
where $\boldsymbol{\rho}_k = \left\{\rho_k \left[n \right] \right\}, \forall n \in \mathcal{N}$ refers to the channel allocation vector for the $k$-th V2V link.

Finally, we design the reward for the BS, which is very crucial to the performance of RL. To maximize the long-term sum rate of V2V links while ensuring the QoS of V2I links in the V2X scenario, we need to devise a mechanism to consider the transmissions of V2V links and V2I links simultaneously. As we know, the V2V links usually carry the safety-critical messages, such as vehicle's speed and emergency vehicle warning on the road, while the V2I links often support the entertainment services \cite{19MARL}. Thus, we should guarantee the transmission of V2V links as the primal target while making sure that the impact of V2V transmission on the V2I links can be tolerable and adjustable to some specific applications. To this end, we model the reward of the BS as
\begin{equation}
R = \lambda_c \sum_{n=1}^N C^c_n\left[n\right]  +  \lambda_d\sum_{k=1}^{K}C^d_k,
\label{Eq9}
\end{equation}
where $C^d_k = \sum_{n=1}^N C^d_k \left[ n \right]$ refers to the capacity of the $k$-th V2V on all the channels. Besides, $\lambda_c$ and $\lambda_d$ are nonnegative weights to balance the performance of V2I links and V2V links.

The solution of the RL problem is related to the concept of policy $\pi \left( a,s\right)$, which defines the probabilities of choosing each action in $\mathcal{A}$ when observing a state in $\mathcal{S}$. The goal of learning is to find an optimal policy $\pi^*$ to maximize the expected return $G_t$ from any initial state $s_0$. The expected return is defined as $G_t = \sum_{k=0}^{\infty} \gamma^k R_{t+k+1}$, which is the cumulative discounted return with a discount factor $\gamma$.

To solve this problem, we resort to the Q-learning \cite{QLearn}, which is a well-known effective approach to tackle the RL problem, due to its model-free property where $p\left(s',r|s, a\right)$ is not required a priori.
Q-learning is based on the idea of action-value function $q_{\pi}\left(s, a\right) = \mathbb{E}_{\pi}\left[G_t |S_t = s, A_t = a \right]$ for a given policy $\pi$, which means the expected return when the agent starts from the state $s$, takes action $a$, and thereafter follows the policy $\pi$. 
The optimal action-value function $q^*\left(s, a\right)$ under the optimal policy $\pi^*$ satisfies the well-known Bellman optimality equations \cite{RLintro}, which can be approached through an iterative update method:
\begin{equation}
 \begin{array}{lr}
{\mathcal{Q}\left(S_t, A_t \right)} \leftarrow {\mathcal{Q}\left(S_t, A_t \right)}
+ \alpha\left[ R_{t+1} + \gamma \max\limits_{a}{\mathcal{Q}\left(S_{t+1}, a\right)}
- {\mathcal{Q}\left(S_t, A_t\right)}\right],
 \end{array}
 \label{Eq10}
\end{equation}
where $\alpha$ is the step-size parameter. Besides, the choice of action $A_t$ in state $S_t$ follows some exploratory policies, such as the $\epsilon$-greedy policy. For better understanding, the $\epsilon$-greedy policy can be expressed as \begin{equation}
 A \leftarrow
 \left\{
 \begin{array}{lr}
   \arg {\max\limits_{a}{\mathcal{Q}\left(s, a\right)}}, \;\;\;\text{ with probability } 1 - \epsilon; \\
   \text{a random action}, \;\;\;\; \text{ with probability } \epsilon.
 \end{array}
 \right.
\label{Eq11}
\end{equation}
Here, $\epsilon$ is also known as the exploration rate in the RL literature. Furthermore, it has been shown in \cite{RLintro} that with a variant of the stochastic approximation conditions on $\alpha$ and the assumption that all the state-action pairs continue to be updated, $\mathcal{Q}$ converges with probability 1 to the optimal action-value function $q^*$.

However, in many practical problems, the state and action space can be extremely large, which prevents storing all action-value functions in a tabular form. As a result, it is common to adopt function approximation to estimate these action-value functions. Moreover, by doing so, we can generalize action-value functions from limited seen state-action pairs to to a much larger space.

In \cite{DQN}, a DNN parameterized by $\boldsymbol{\theta}$ is employed to represent the action-value function, thus called as DQN. 
DQN adopts the $\epsilon$-greedy policy to explore the state space and store the transition tuple $\left( S_t, A_t,R_{t+1}, S_{t+1} \right)$ in a replay memory (also known as the replay buffer) at each time step. The replay memory accumulates agent's experiences over many episodes of the MDP. At each time step, a mini-batch of experiences $\mathcal{D}$ are uniformly sampled from the replay memory,  called experience replay, to update the network parameters $\boldsymbol{\theta}$ with variants of stochastic gradient descent method to minimize the squared errors shown as follows:
\begin{equation}
\sum_{t \in \mathcal{D}}\left[R_{t+1} + \gamma \max\limits_{a}{\mathcal{Q}\left(S_{t+1}, a; \boldsymbol{\theta}^{-}\right)} - {\mathcal{Q}\left(S_t, A_t; \boldsymbol{\theta} \right)} \right]^2,
\label{Eq12}
\end{equation}
where $\boldsymbol{\theta}^{-}$ is the parameter set of a target Q-network, which is duplicated from the training Q-network parameter set $\boldsymbol{\theta}$, and fixed for a couple of updates with the aim of further improving the stability of DQN. Besides, experience replay improves sample efficiency via repeatedly sampling experiences from the replay memory and also breaks correlation in successive updates, which also stabilizes the learning process.


\subsection{Centralized Control and Distributed Transmission Architecture}

\begin{algorithm}
\caption{Training algorithm for the C-Decision scheme}
\hspace*{0.02in}{\bf Input:}
the DNN model for each V2V, the DQN model for the BS, the V2X environment 
\hspace*{10mm}simulator \\
\hspace*{0.02in}{\bf Output:}
the DNN for each V2V, the optimal control policy $\pi^*$ represented by a DQN $\mathcal{Q}$ with 
\hspace*{10mm}parameters $\boldsymbol{\theta}$
\begin{algorithmic}[1]
    \State Initialize all DNNs and DQN models respectively
\For {episode $l = 1, ..., L_{train}$}
    \State Start the V2X environment simulator, generate vehicles, V2V links and V2I links
    \State Initialize the beginning observations $\mathbf{O}_0$
    \State Initialize the policy $\pi$ randomly
    \For {time-step $t = 1,...,T$}
        \State Each V2V adopts the current observation $\mathbf{o}_k^t$ as the input of its DNN to learn the  
        \item[]\hspace{\algorithmicindent} \hspace{\algorithmicindent}feedback $\mathbf{b}_k^t$, and  sends it to BS
        \State BS takes $\mathbf{s}_t = \left\{ \mathbf{b}_k^t \right\}$ as the input of its DQN $\mathcal{Q}$
        \State BS chooses $\mathbf{a}_t$ according to $\mathbf{s}_t$ using some policy $\pi$ derived from $\mathcal{Q}$, e.g., $\epsilon$-greedy
        \item[]\hspace{\algorithmicindent} \hspace{\algorithmicindent}strategy as in (\ref{Eq11})
        \State BS broadcasts the action $\mathbf{a}_t$ to every V2V
        \State Each V2V takes action based on $\mathbf{a}_t$, and gets the reward $R_{t+1}$ and the next observation         \item[]  \hspace{\algorithmicindent} \hspace{\algorithmicindent}$\mathbf{o}_k^{t+1}$
        \State Save the data $\left\{\mathbf{O}_t, \mathbf{a}_t, R_{t+1},  \mathbf{O}_{t+1} \right\}$ into the replay buffer $\mathcal{B}$
        \State Sample a mini-batch of data $\mathcal{D}$ from $\mathcal{B}$ uniformly
        \State Use the data in $\mathcal{D}$ to train all V2Vs' DNNs and BS's DQN together as in (\ref{Eq14})
        \State Each V2V updates its observation $\mathbf{o}_k^t \leftarrow \mathbf{o}_k^{t+1}$
        \State Update the target network: $\boldsymbol{\theta}^{-}\leftarrow \boldsymbol{\theta}$ every $N_{u}$ steps
    \EndFor
\EndFor
\end{algorithmic}
\label{Algo1}
\end{algorithm}

In this part, the architecture for the C-Decision scheme is shown in Fig. \ref{Fig1}. Each V2V link first observes its local environment and then adopts a DNN to compress the observed information into several real numbers, which are finally fed back to the BS for centralized decision making. 
The BS takes the feedback information of all V2V links as the input, utilizes the DQN to perform Q-learning to decide the channel allocation for all V2V links, and broadcasts its decision. 
Finally, each V2V link chooses the BS-allocated channel for its transmission. 

Details of the training framework for the C-Decision scheme are provided in Algorithm \ref{Algo1}. We define $\mathbf{O}_t = \left\{ \mathbf{o}_k^t \right\}, \forall k \in \mathcal{K}$ as the observations of all V2Vs at the time step $t \in \left\{1,2,..., T \right\}$, where $\mathbf{o}_k^t$ refers to the observation of the $k$-th V2V at the time step $t$.  Then, we can express the estimation of the return also known as the approximate target value \cite{DQN}  as
\begin{equation}
y_t = R_{t+1} + \gamma \max\limits_{\mathbf{a}}{\mathcal{Q}\left(\mathbf{O}_{t+1}, \mathbf{a}; \boldsymbol{\theta}^{-}\right)},
\label{Eq13}
\end{equation}
where $R_{t+1}$ and ${\mathcal{Q}\left(\mathbf{O}_{t+1}, \mathbf{a}; \boldsymbol{\theta}^{-}\right)}$ represent the reward of all links and the Q function of the target DQN with parameters $\boldsymbol{\theta}^{-}$ under the next observation $\mathbf{O}_{t+1}$ and the action $\mathbf{a}$, respectively. Then, the updating process for the BS DQN can be written as \cite{DQN,16DoubleQN}:
\begin{equation}
\boldsymbol{\theta} \leftarrow \boldsymbol{\theta} + \beta \sum_{t \in \mathcal{D}} \frac{\partial {\mathcal{Q}\left(\mathbf{O}_t, \mathbf{a}_t; \boldsymbol{\theta} \right)}}{ \partial {\boldsymbol{\theta}}}\left[y_t - {\mathcal{Q}\left(\mathbf{O}_t, \mathbf{a}_t; \boldsymbol{\theta} \right)}\right] ,
\label{Eq14}
\end{equation}
where $\beta$ is the step size in one gradient iteration. 

As for the testing phase, at each time step $t$, each V2V adopts its observation $\mathbf{o}_k^t$ as the input of the trained DNN to obtain its learned feedback $\mathbf{b}_k^t$, and then sends it to the BS. After that, the BS takes $\left\{ \mathbf{b}_k^t \right\}$ as the input of its trained DQN to generate the decision result $\mathbf{a}_t$, and broadcasts $\mathbf{a}_t$ to all V2Vs. Finally, each V2V chooses the specific channel indicated by $\mathbf{a}_t$ to transmit.

\subsection{Spectrum Sharing with Binary Feedback}

In order to further reduce feedback overhead, we propose a framework to quantize the V2V links' real-valued feedback into several binary digits. In other words, we try to constrain ${b_{k,j}} \in \left\{-1,1 \right\},\forall k \in \mathcal{K}, \forall j \in \left\{1,2,..., N_k \right\}$. 
The binarization procedure can help force the neural networks to learn efficient representations of the feedback information compared to the standard floating-point layer. In other words, a binary layer can make each V2V compress its observation more efficiently.

The binary quantization process consists of two steps \cite{binaryref}. The first step is to generate the required number of continuous feedback values in the continuous interval $\left[-1,1 \right]$, which is also equal to the desired number of the binary feedback. Then, the second step takes the outputs of the first step as its input to produce the desired number of discrete feedback in the set $\left\{-1,1 \right\}$ for each output real-valued feedback of the first step. 

For the first step, we adopt a fully-connected layer with $\tanh$ activations, defined as $\tanh\left( x\right) = \frac{2}{1 + e^{-2x}} - 1$, where we term this layer as the pre-binary layer. Here, the input of this pre-binary layer connects the outputs of each V2V's DNN. 
Then, in order to binarize the continuous output of the first step, we adopt the traditional sign function method in the second step. To be specific, we take the sign of the input value as the output of this layer, which is shown as below:
\begin{equation}
b\left(x\right) = \left\{\begin{aligned} 1,  &&  x \ge 0; \\ -1, && x <0.  \end{aligned} \right.
\end{equation}
However, the gradient of this function is not continuous, challenging the back propagation procedure for DNN training. As a remedy to this, we adopt the identity function in the backward pass, which is known as the straight-through estimator \cite{bengio2013estimating}.
Combining these two steps together, we can express the full binary feedback function as
\begin{equation}
B\left( x \right) = b\left(\tanh\left(W_{0}x + b_{0}\right)\right),
\label{Eq16}
\end{equation}
where $W_0$ and $b_0$ denote the linear weights and bias of the pre-binary layer that transform the activations from the previous layer in the neural network respectively. Here, we term this layer as the binary layer.

Finally, to implement the C-Decision scheme with binary feedback, we add the full binary feedback function in (\ref{Eq16}), which consists of the pre-binary layer in the first step and the binary layer in the second step, to the output of each V2V link's DNN. Besides, in response to the change in the number of feedback bits at each V2V link's new DNN, the number of inputs in the DQN of BS should change correspondingly.

\section{Distributed Decision Making and Spectrum Sharing Architecture}
\begin{figure}
\begin{center}
\includegraphics[width=0.6\textwidth]{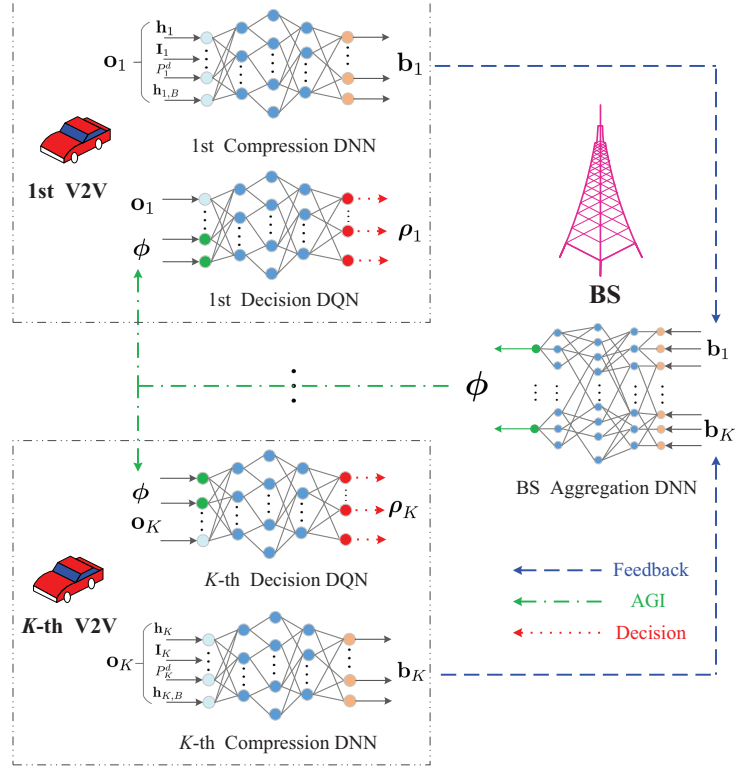}
\caption{Neural network architecture for V2V links and the BS in the D-Decision scheme.}
\label{Fig101}
\end{center}
\end{figure}

In order to further facilitate distributed spectrum sharing and reduce the computational complexity, we propose the distributed decision making and spectrum sharing architecture (named as the D-Decision scheme) shown in Fig. \ref{Fig101} to let each V2V link make its own spectrum sharing decision. In this section, we first devise the neural network architecture for each V2V link to compress CSI and make decision, respectively, and then design the neural network for the BS to aggregate feedback from all V2V links. Then, we propose the hybrid information aggregation and distributed control architecture. Finally, we propose the D-Decision scheme with the binary aggregated information.   

\subsection{DNN Design at V2V and BS}
To enable distributed decision making, each V2V contains one DNN to compress local observations for feedback, termed the Compression DNN and another DQN for distributed spectrum sharing decision making, termed Decision DQN. 
Here, we employ the same DNN architecture for each V2V as that in Part A of Section III since they share the same functionality.  

The BS aggregates the feedback from all V2Vs via its DNN, termed as the Aggregation DNN, and then broadcasts the aggregated global information (AGI) $\boldsymbol{\phi}$ to all V2Vs. Here, the AGI can be expressed as $\boldsymbol{\phi} = \left\{ \phi_j \right\}, \forall j \in \left\{1,2,...,N_g\right\}$, where $N_g$ refers to the number of AGI values and also equals the number of outputs of BS Aggregation DNN. Finally, each V2V combines its local observation and the AGI as the input of its Decision DQN to decide which channel to transmit. 

\subsection{Hybrid Information Aggregation and Distributed Control Architecture}
Each V2V link first observes its local environment to obtain $\mathbf{o}_k$, and then adopts its Compression DNN to compress $\mathbf{o}_k$ into several real numbers $\mathbf{b}_k$, and finally feeds this compressed information back to the BS. 
After that, the BS takes the feedback values of all V2V links $\left\{ \mathbf{b}_k \right\}$ as the input of its Aggregation DNN to aggregate the compressed observations of all V2V links and further compress this information into the AGI $\boldsymbol{\phi}$.
Finally, each V2V link combines the received AGI $\boldsymbol{\phi}$ and its local observation $\mathbf{o}_k$ as the input of its Decision DQN, and performs the Q-learning algorithm to decide which channel to transmit. 

Details of the training framework for the D-Decision scheme are provided in Algorithm \ref{Algo3}. Here, we define $\mathbf{a}_t = \left\{ \mathbf{a}_k^t \right\}, \forall k \in \mathcal{K}$ as the actions of all V2V links at the time step $t \in \left\{1,2,..., T \right\}$, where $\mathbf{a}_k^t = \boldsymbol{\rho}_k$ refers to the action for $k$-th V2V. Besides, in the training process, we take the observations of all V2V links $\mathbf{O}_t$ as the input and train all DNNs and DQNs in an end-to-end manner. 
The training process can be implemented in a fully distributed manner.

As for the testing phase, at each time step $t$, each V2V link adopts its observation $\mathbf{o}_k^t$ as the input of its Compression DNN to learn the feedback $\mathbf{b}_k^t$, and sends it to the BS.
Then, the BS utilizes  $\left\{ \mathbf{b}_k^t \right\}$ as the input of its Aggregation DNN to generate the AGI $\boldsymbol{\phi}_t$, and broadcasts  $\boldsymbol{\phi}_t$ to all V2V links.
Finally, each V2V link takes $\mathbf{o}_k^t$ and $\boldsymbol{\phi}_t$ as the input of its Decision DQN to make decision, and then transmits on the chosen channel.

\begin{algorithm}
\caption{Training algorithm for the D-Decision scheme}
\hspace*{0.02in}{\bf Input:}
the Compression DNN and Decision DQN for each V2V, the Aggregation DNN for the  \hspace*{10mm}BS, the V2X environment simulator \\
\hspace*{0.02in}{\bf Output:}
the Compression DNN, the optimal policy $\pi_k^*$ represented by the Decision DQN $\mathcal{Q}_k$ 
\hspace*{10mm}with parameters $\boldsymbol{\theta}_k$ for each V2V, the Aggregation DNN for the BS
\begin{algorithmic}[1]
    \State Initialize all DNNs and DQNs models respectively
\For {episode $l = 1, ..., L_{train}$}
    \State Start the V2X environment simulator, generate vehicles, V2V links and V2I links
    \State Initialize the beginning observations $\mathbf{O}_0$
    \State Initialize the policy $\pi_k$ for each V2V randomly
    \For {time-step $t = 1,...,T$}
        \State Each V2V adopts the current observation $\mathbf{o}_k^t$ as the input of its Compression DNN  
        \item[]\hspace{\algorithmicindent} \hspace{\algorithmicindent}to learn the feedback $\mathbf{b}_k^t$, and sends it to BS
        \State BS takes $\left\{ \mathbf{b}_k^t \right\}$ as the input of its Aggregation DNN, and generates the AGI $\boldsymbol{\phi}_t$
        \State BS broadcasts the AGI $\boldsymbol{\phi}_t$ to every V2V
        \State Each V2V takes $\mathbf{s}_k^t = \left\{ \mathbf{o}_k^t, \boldsymbol{\phi}_t \right\}$ as the input of its Decision DQN $\mathcal{Q}_k$
        \State Each V2V chooses $\mathbf{a}_k^t$ according to $\mathbf{s}_k^t$ using some policy $\pi_k$ derived from $\mathcal{Q}_k$, e.g.,  
        \item[]\hspace{\algorithmicindent} \hspace{\algorithmicindent}$\epsilon$-greedy strategy as in (\ref{Eq11}) 
        \State Each V2V takes action $\mathbf{a}_k^t$, and gets the reward $R_{t+1}$ and the next observation $\mathbf{o}_k^{t+1}$
        \State Save the data $\left\{\mathbf{O}_t, \mathbf{a}_t, R_{t+1},  \mathbf{O}_{t+1} \right\}$ into the buffer $\mathcal{B}$
        \State Sample a mini-batch of data $\mathcal{D}$ from $\mathcal{B}$ uniformly
        \State Use the data in $\mathcal{D}$  to train all V2Vs' Compression DNNs and Decision DQNs  
        \item[]  \hspace{\algorithmicindent} \hspace{\algorithmicindent}and BS's Aggregation DNN together as in (\ref{Eq14})
        \State Each V2V updates its observation $\mathbf{o}_k^t \leftarrow \mathbf{o}_k^{t+1}$
        \State Each V2V updates its target network: $\boldsymbol{\theta}_k^{-}\leftarrow \boldsymbol{\theta}_k$ every $N_{u}$ steps
    \EndFor
\EndFor
\end{algorithmic}
\label{Algo3}
\end{algorithm}

\subsection{Distributed Spectrum Sharing with binary information}
Similar to Section III-D, we can also quantize the continuous feedback and the AGI in the D-Decision scheme into the binary data to further reduce signaling overhead. Then, both the Compression DNN of each V2V link and the Aggregation DNN at the BS need to include the binary function in (\ref{Eq16}).

\section{Simulation Results}
In this section, we conduct extensive simulation to verify the performance of the proposed schemes. In particular, we provide the simulation settings in Part A, and evaluate the training performance of the C-Decision scheme in Part B. Then, we assess the testing performance under the real-valued feedback and binary feedback in Parts C and D respectively. Besides, we demonstrate the impacts of V2I and V2V links weights on the performance in Part E and the robustness of the proposed scheme in Part F, respectively. Finally, we show the training and testing performance of the D-Decision scheme in Part G.  

\subsection{Simulation Settings}
The simulation scenario follows the urban case in Annex A of \cite{SimScenario}. The simulation area size is $1,299 \; m \times 750 \; m$, where the BS is located in the center of this area. For better understanding, we provide related parameters and their corresponding settings in Table \ref{TB1}. In addition, we list the corresponding channel models for both V2V and V2I links respectively in Table \ref{TB2}.

\begin{table}[h]
\centering
\caption{Simulation Parameters}
\begin{tabular}{|l|c|}
\hline
Parameters&Typical values\\
\hline
Number of V2I links $N$ & 4 \\
\hline
Number of V2V links $K$ & 4 \\
\hline
Carrier frequency & 2 GHz \\
\hline
Normalized Channel Bandwidth  & 1 \\
\hline
BS antenna height  & 25 m\\
\hline
BS antenna gain  & 8 dBi\\
\hline
BS receive noise figure  & 5 dB\\
\hline
Vehicle antenna height  & 1.5 m\\
\hline
Vehicle antenna gain  & 3 dBi\\
\hline
Vehicle receive noise figure  & 9 dB\\
\hline
Vehicle speed &  randomly in [10, 15] km/h\\
\hline
Vehicle drop and mobility model &  Urban case of A.1.2 in \cite{SimScenario}\\
\hline
V2I transmit power $P_{n}^c$& 23 dBm\\
\hline
V2V transmit power $P_{k}^d$& 10 dBm\\
\hline
\end{tabular}
\label{TB1}
\end{table}

\begin{table*}[!t]
\centering
\caption{Channel models for V2I and V2V links}
\resizebox{0.8\textwidth}{!}{
\begin{tabular}{|l|l|l|}
\hline
Parameter & V2I link & V2V link \\
\hline
Path loss model  & $128.1 + 37.6\log_{10}\left(d\right)$, d in km & LOS in WINNER + B1 Manhattan \cite{WinnerII}\\
\hline
Shadowing distribution & Log-normal & Log-normal\\
\hline
Shadowing standard deviation  & 8 dB & 3 dB \\
\hline
Decorrelation distance  & 50 m & 10 m \\
\hline
Noise power $\sigma^2$ & -114 dBm  &  -114 dBm \\
\hline
Fast fading & Rayleigh fading & Rayleigh fading \\
\hline
Fast fading update & Every 1 ms & Every 1 ms \\
\hline
\end{tabular}}
\label{TB2}
\end{table*}

The specific architecture of DNNs and BS DQN under the C-Decision scheme are summarized in Table \ref{TB3}, where $N_k$ to refers the number of feedback for each V2V link and FC denotes the fully connected (FC) layer respectively. In addition, the number of neurons in the output layer of the BS DQN is set as $256$, which refers to all the possible channel allocations for all V2V links under current simulation setting. Besides, the settings for the DNNs and DQNs under the D-Decision scheme are listed in Table \ref{TB4}.

\begin{table}  
\centering
\caption{Architecture for DNN and BS DQN in the C-Decision scheme}
\begin{tabular}{|l|l|l|}
\hline
     & DNN & BS DQN \\
\hline
Input layer  &  13 & $K \times N_k$ \\
\hline
Hidden layers & 3 FC layers (16, 32, 16)  & 3 FC layers (1200, 800, 600) \\
\hline
Output layer  & $N_k$ &  $256$ \\
\hline
\end{tabular}
\label{TB3}
\end{table}

\begin{table}  
\centering
\caption{Architecture for DNNs and DQNs in the D-Decision scheme}
\begin{tabular}{|l|l|l|l|}
\hline
     & Compression DNN & Aggregation DNN  & Decision DQN \\
\hline
Input layer  &  13 & $K \times N_k$ &   $N_g$\\
\hline
Hidden layers & 3 FC layers (16, 32, 16)  &  3 FC layers (500, 400, 300)  &  3 FC layers (80, 40, 20) \\
\hline
Output layer  & $N_k$ &  $N_g$ &   $4$  \\
\hline
\end{tabular}
\label{TB4}
\end{table}

We use the rectified linear unit (ReLU) activation function for both DNN and DQNs, defined as $f\left(x\right) = \max{\left(0, x\right)}$. Here, the activation function of output layers in DNNs and DQNs is set as a linear function. Besides, the RMSProp optimizer \cite{DLOptimizer} is adopted to update the network parameters with a learning rate of $0.001$. The loss function is set as the Huber loss \cite{HuberLoss}.

We choose the weights $\lambda_c = 0.1$ and $\lambda_d = 1$ for V2I and V2V links, respectively. We train the whole neural network for $2,000$ episodes and the exploration rate $\epsilon$ is linearly annealed from $1$ to $0.01$ over the beginning $1,600$ episodes and keeps constant afterwards. The number of steps in each episode is set as $T=1,000$. The update frequency $N_u$ of the target Q-network is every $500$ steps. The discount factor, $\gamma$, in the training is chosen as $0.05$. The size of the replay buffer $\mathcal{B}$ is set as $1,000,000$ samples. Meanwhile, the mini-batch size $\mathcal{D}$ varies in different settings, to be specified in each figure.

\subsection{Training Performance Evaluation}

\begin{figure} \centering
\subfigure[Training loss] { \label{fig3:a}
\includegraphics[width=0.45\textwidth]{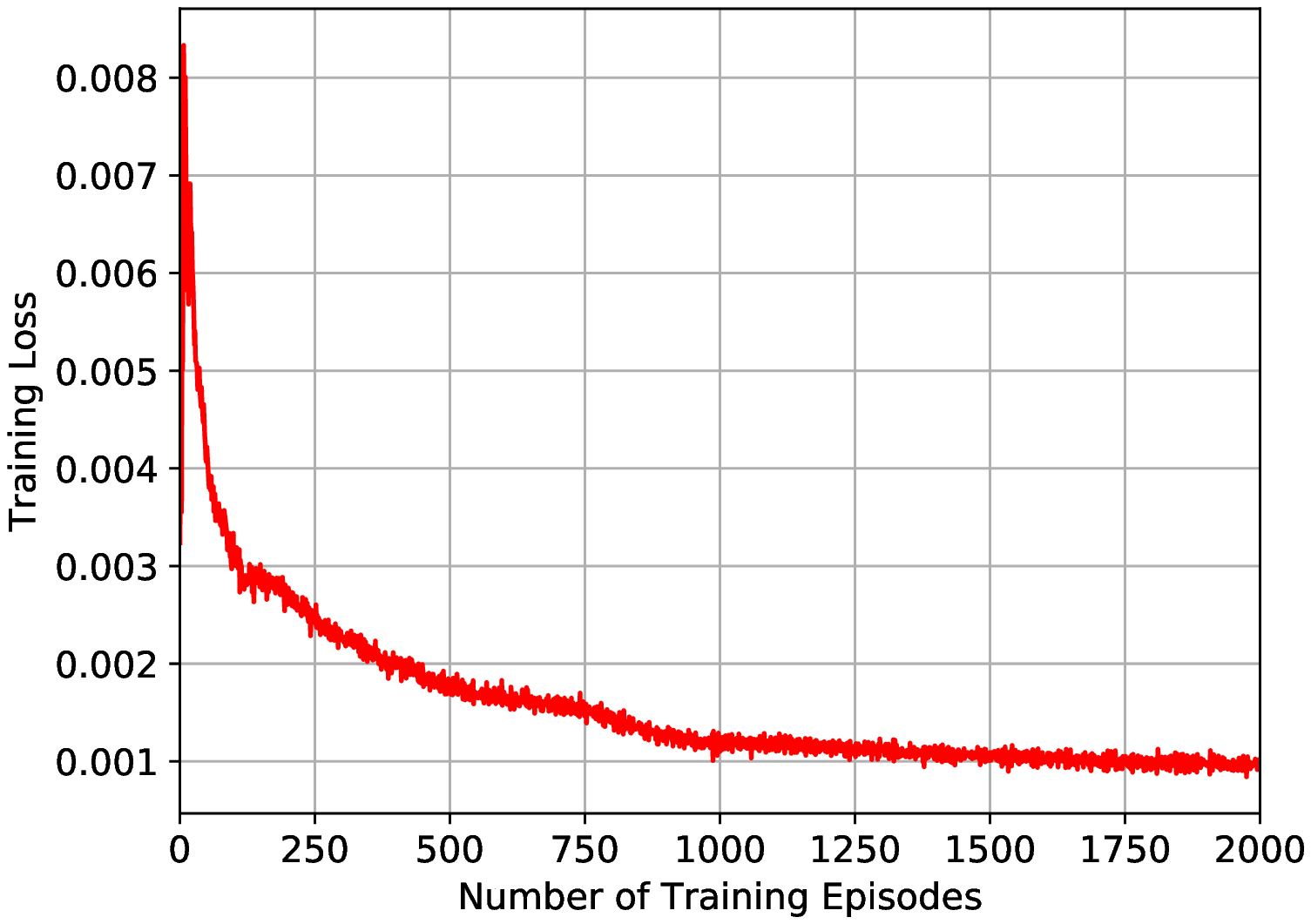}
}
\subfigure[Average return per episode] { \label{fig3:b}
\includegraphics[width=0.45\textwidth]{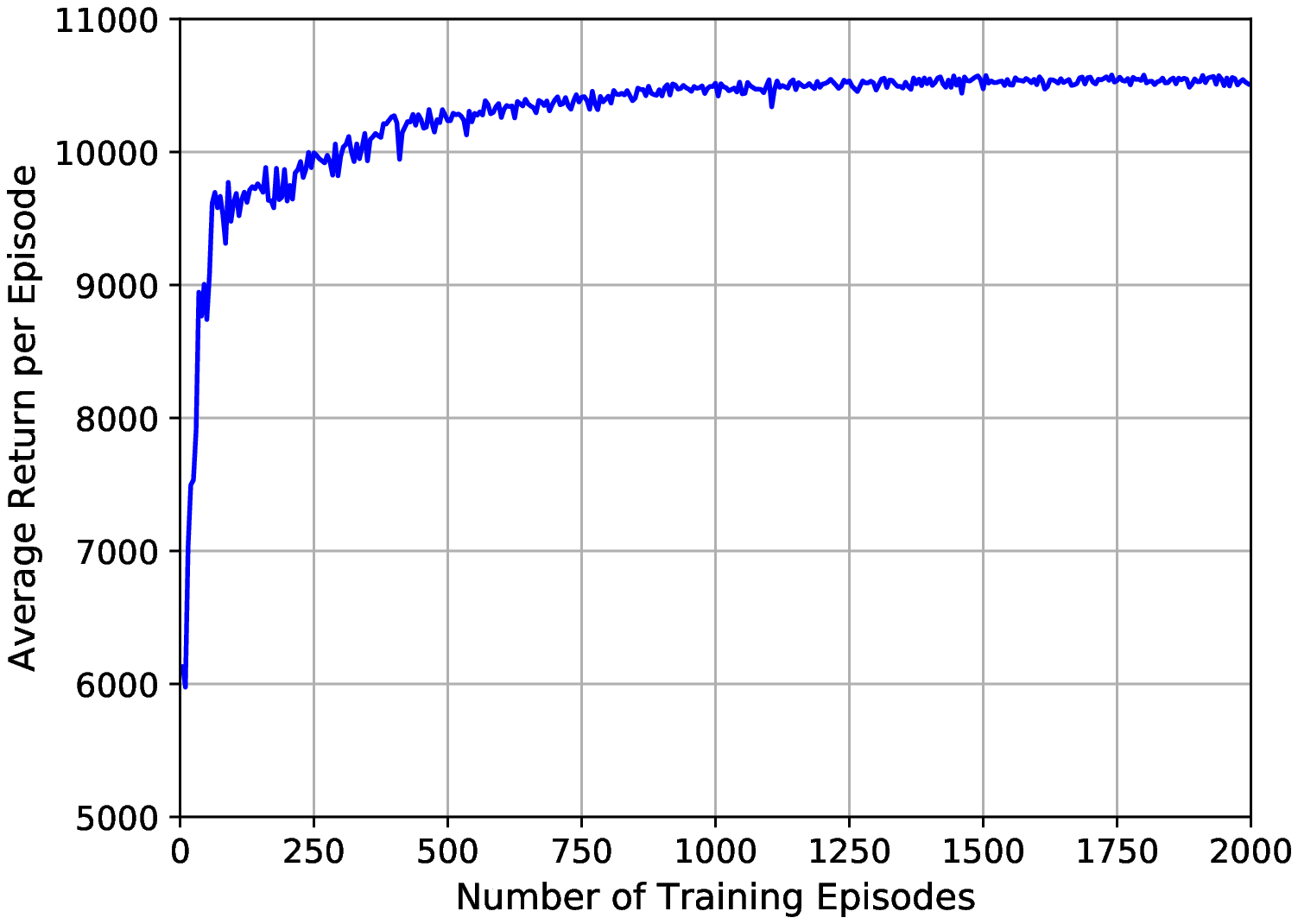}

}
\caption{Training performance evaluation for the C-Decision scheme.}
\label{fig3}
\end{figure}

Fig. \ref{fig3} demonstrates the training performance of the proposed C-Decision scheme with a mini-batch size $\mathcal{D} = 512$ and the number of real-valued feedback $N_k = 3$. In Fig. \ref{fig3} \subref{fig3:a}, the loss function decreases quickly with the increasing number of training episodes $L_{train}$, and becomes nearly unchanged with the further increasing $L_{train}$. On the other hand, the change of average return per episode is displayed in Fig. \ref{fig3} \subref{fig3:b}. Here, we evaluate the training process every $5$ training episodes under $10$ different random seeds with the exploration rate $\epsilon=0$, and plot the average return per episode in Fig. \ref{fig3} \subref{fig3:b}. The average return per episode first increases quickly with increasing $L_{train}$, and gradually converges despite some small fluctuations due to the time-varying V2X scenario, which shows the stability of the training process. Thus, Fig. \ref{fig3} \subref{fig3:a} and \subref{fig3:b} demonstrate the desired convergence of the proposed training algorithm. Therefore, we set $L_{train} = 2,000$ for the C-Decision scheme afterwards.


\subsection{Performance of Real-Valued Feedback }
\begin{figure} \centering
\subfigure[Normalized return comparison] { \label{Fig4:a}
\includegraphics[width=0.45\textwidth]{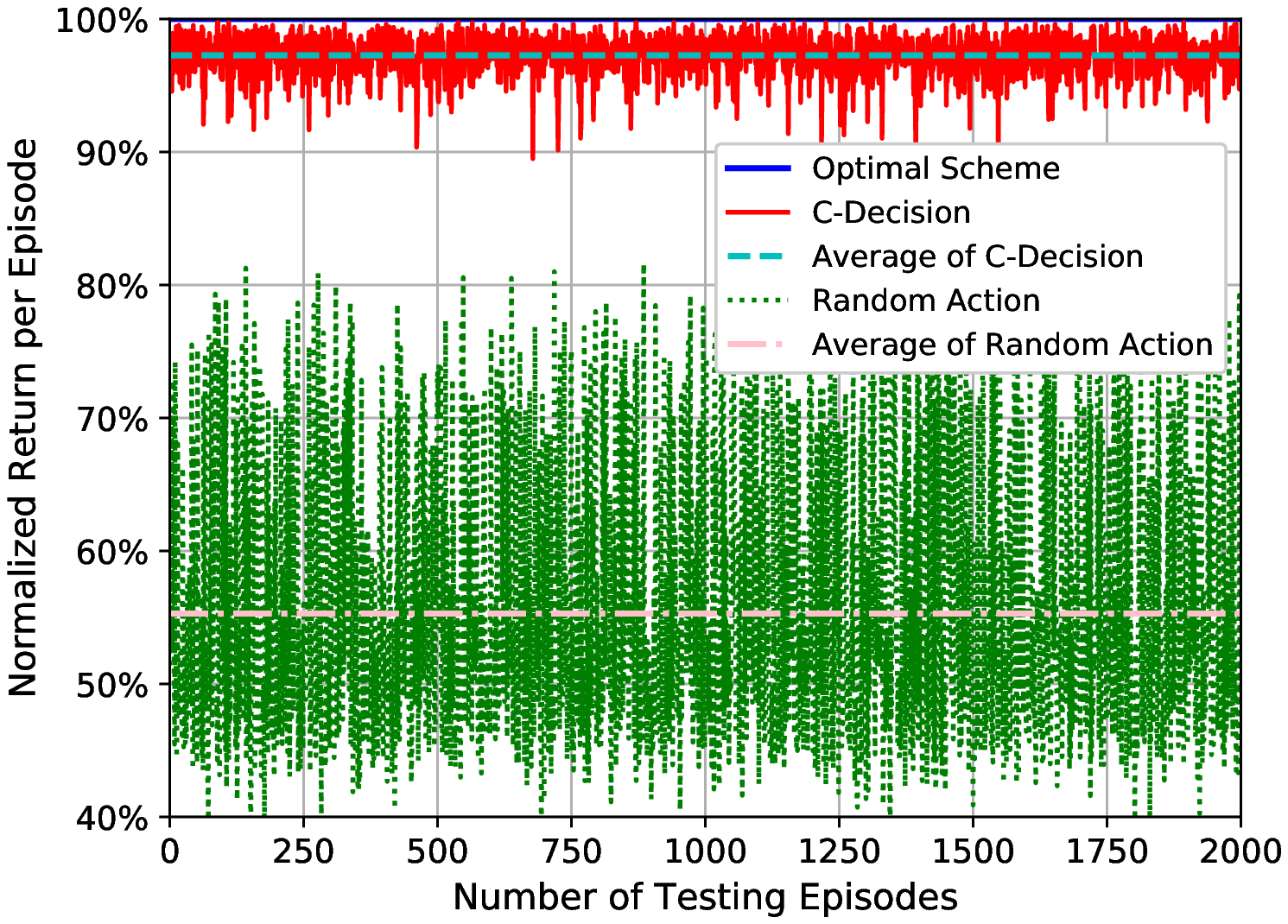}
}
\subfigure[ARP performance with real-valued feedback] { \label{Fig4:b}
\includegraphics[width=0.45\textwidth]{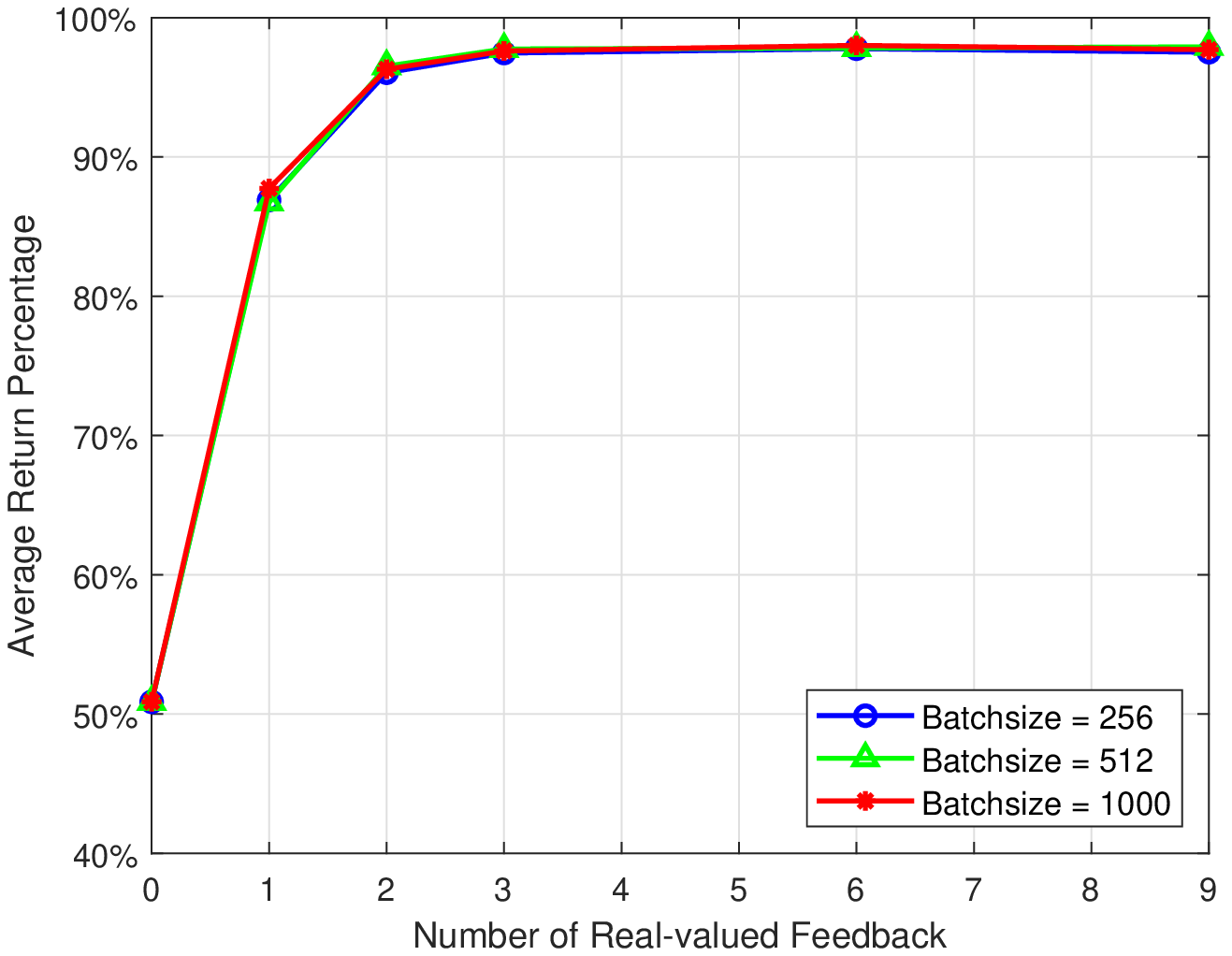}
}
\caption{Performance evaluation for the C-Decision scheme with real-valued feedback.}
\label{Fig4}
\end{figure}

Fig. \ref{Fig4} \subref{Fig4:a} shows the return variation under the real-valued feedback with the number of testing episodes. Here, we choose the mini-batch size as $\mathcal{D}=512$, number of testing episodes as $L_{test} = 2,000$, and the number of real-valued feedback as $N_k = 3$, respectively. For comparison, we also display the performance of two benchmark schemes: the optimal and the random action schemes, respectively. In the optimal scheme, we perform time-consuming brute-force search to find the optimal spectrum allocation in each testing step. In the random action scheme, each V2V link chooses the channel randomly. For better comparison, we depict the normalized return of these three schemes in Fig. \ref{Fig4} \subref{Fig4:a}, where we use the return of the optimal scheme to normalize the return of the other two schemes in each testing episode. Besides, the average return of our proposed scheme and the random action scheme are also depicted. In Fig. \ref{Fig4} \subref{Fig4:a}, the performance of the C-Decision approaches $100\%$ in most episodes and its average performance is about $97\%$ of the optimal scheme while the average performance of random selection is about $55\%$ of the optimal performance. Thus, we conclude the proposed C-Decision scheme can achieve near-optimal spectrum sharing.

Fig. \ref{Fig4} \subref{Fig4:b} shows the impacts of different mini-batch sizes $\mathcal{D}$ and different numbers of real-valued feedback $N_k$ on the performance of the C-Decision scheme, which adopts the average return percentage (ARP) as the metric. Here, the ARP metric is defined as: the return under the C-Decision scheme is first averaged over $2,000$ testing episodes and then normalized by the average return of the optimal scheme. In Fig. \ref{Fig4} \subref{Fig4:b}, the number of real-valued feedback equals $0$ refers to the situation where each V2V link does not feed anything back to the BS and therefore, each V2V link just randomly selects channel to transmit, which is known as the random action scheme. From Fig. \ref{Fig4} \subref{Fig4:b}, the ARP under the C-Decision scheme increases rapidly with the increase of $N_k$, and reaches the maximal percentage nearly $98\%$ at $N_k = 3$. Thereafter, the ARP virtually keeps constant with increasing $N_k$. In other words, each V2V link only needs to send $3$ real-valued feedback to the BS to achieve near-optimal performance. Besides, different mini-batch sizes can achieve very similar performance. Particularly, the mini-batch size $\mathcal{D} = 512$ achieves the best performance, which is good enough considering the computational overhead in the training process and the gained performance.

\subsection{Performance of Binary Feedback}
\begin{figure}
\begin{center}
\includegraphics[width=0.45\textwidth]{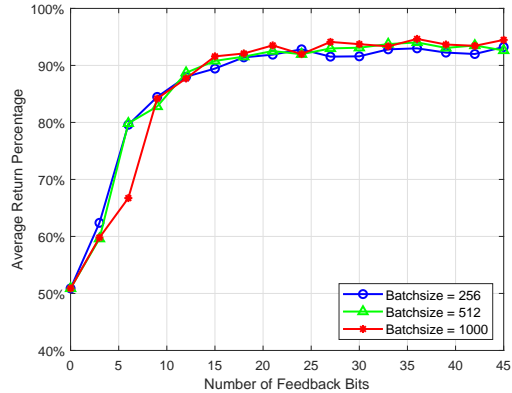}
\caption{The ARP performance for the C-Decision scheme with binary feedback.}
\label{Fig6}
\end{center}
\end{figure}

Fig. \ref{Fig6} demonstrates the change of the ARP performance with an increasing number of feedback bits under different mini-batch sizes. Here, we fix the number of real-valued feedback as $3$, and quantize these real-valued feedback into different numbers of feedback bits.  Similarly, the number of feedback bits equals $0$ in Fig. \ref{Fig6} refers to the situation where each V2V link does not feedback anything to the BS and just adopts the random action scheme. 
The ARP first increases quickly with the number of feedback bits, and then keeps nearly unchanged with the further increasing of feedback bits after the number of feedback bits is larger than $21$. The ARP under different $\mathcal{D}$ has quite similar performance. Besides, the ARP can reach $94\%$ with $36$ feedback bits under $\mathcal{D}=512$. Considering the performance and feedback overhead tradeoff, we choose $36$ feedback bits under $\mathcal{D}=512$ in the subsequent evaluation.

\subsection{Impacts of V2I and V2V Weights}
\begin{figure} \centering
\subfigure[V2I sum rate comparison] {\label{fig11:a}
\includegraphics[width=0.45\textwidth]{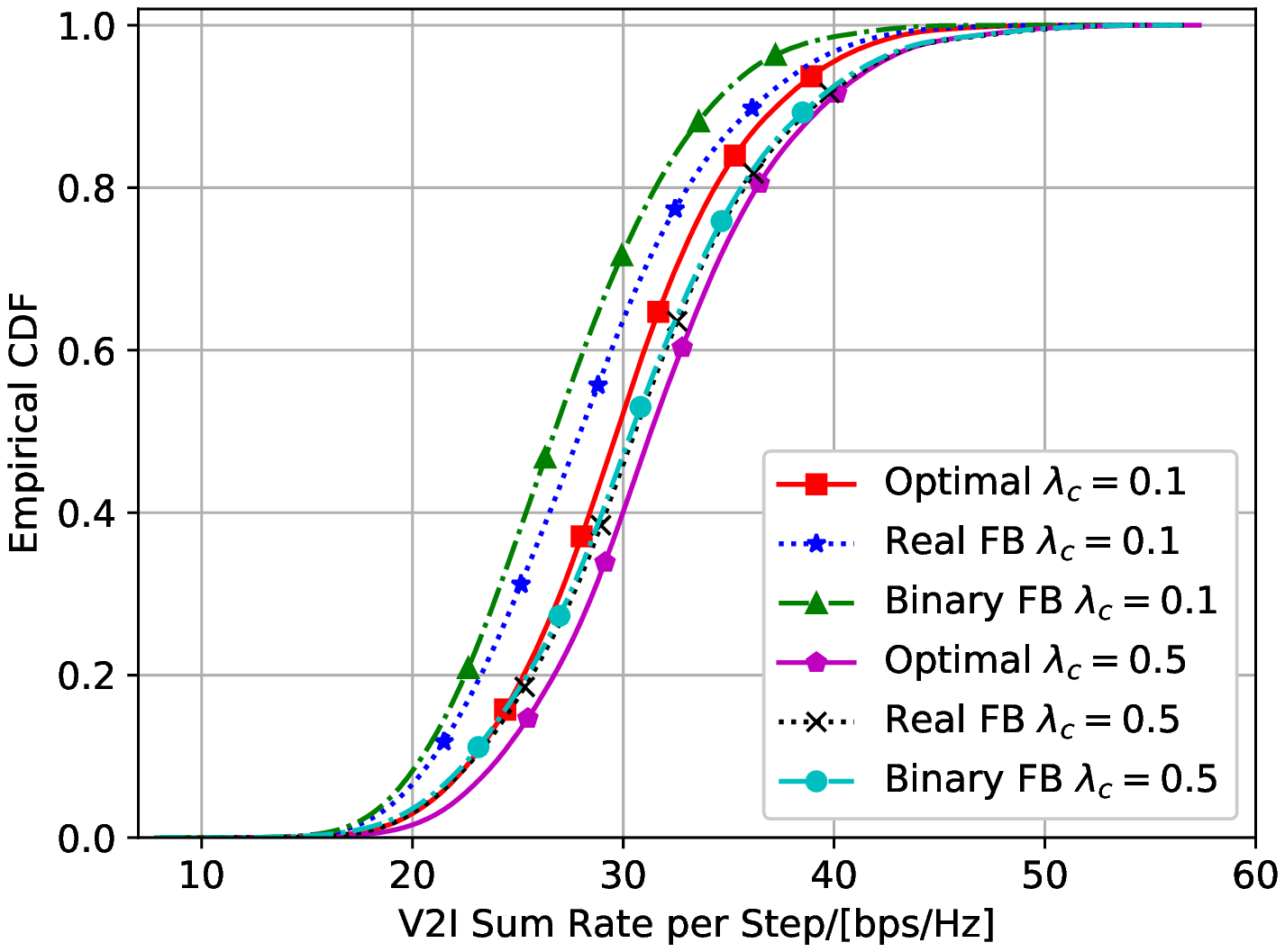}
}
\subfigure[V2V sum rate comparison (same legend as Fig. \ref{fig11} \subref{fig11:a})] {\label{fig11:b}
\includegraphics[width=0.45\textwidth]{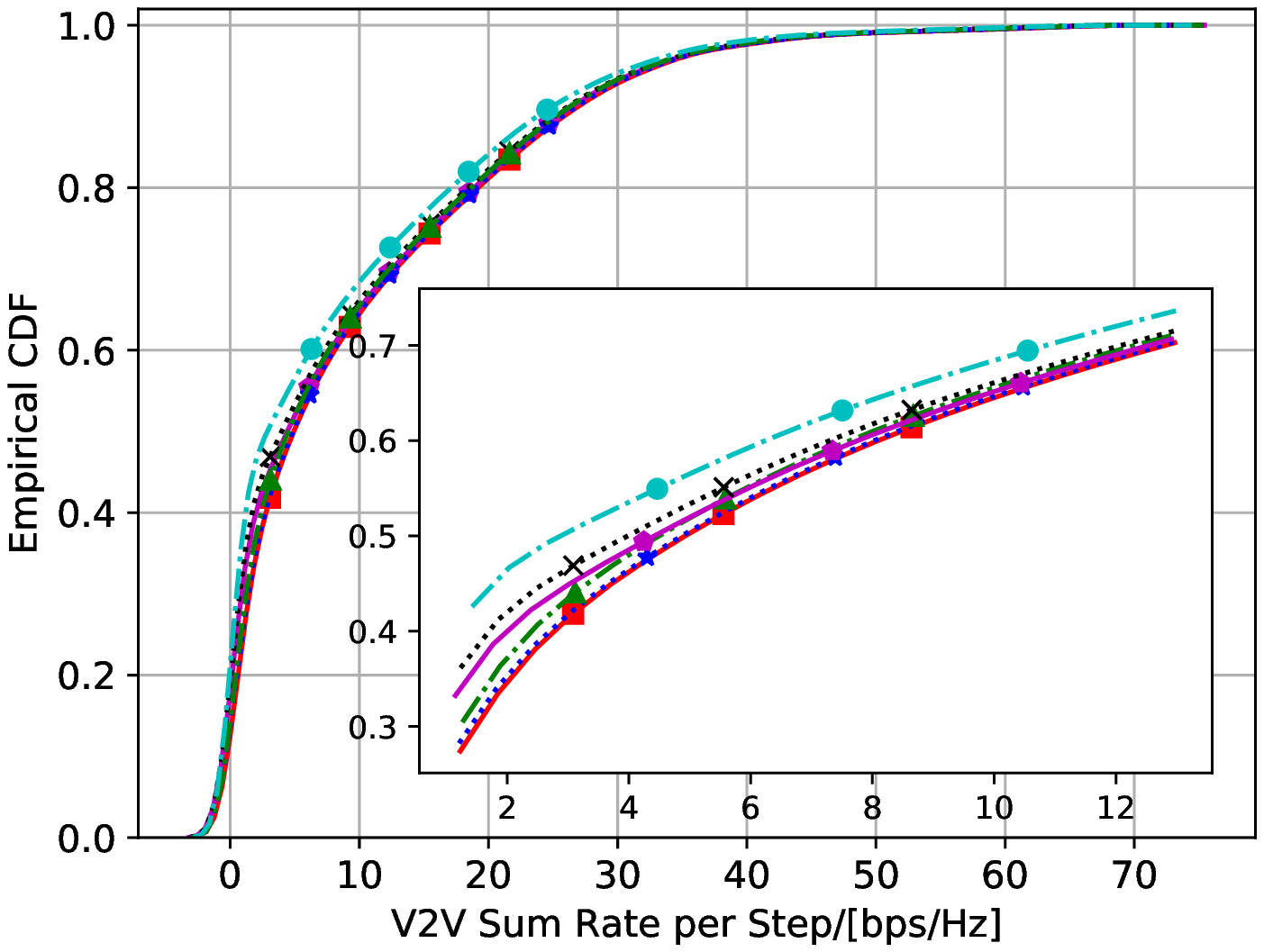}
}
\caption{Sum rate performance with different weights.}
\label{fig11}
\end{figure}

In this part, we evaluate the impacts of V2I links weight $\lambda_c$ and V2V links weights $\lambda_d$ on the system performance. For better understanding, we fix $\lambda_d = 1$ and vary the values of $\lambda_c$. Fig. \ref{fig11} demonstrates the empirical cumulative distribution function (CDF) of V2I and V2V sum rate. In Fig. \ref{fig11}, ``Real FB'' and ``Binary FB'' refer to the proposed C-Decision scheme with real-valued feedback and that with binary feedback respectively, and ``Optimal'' represents the optimal scheme. In particular, two empirical CDFs of V2I sum rate under both real-valued feedback and binary feedback in Fig. \ref{fig11} \subref{fig11:a} shift quickly to the right when the V2I weight $\lambda_c = 0.1$ increases to $0.5$, which shows our proposed scheme can ensure different QoS requirements of V2I links via adjusting $\lambda_c$. Besides, the performance gap between the real-valued feedback and binary feedback decreases with the increase of $\lambda_c$. From Fig.~ \ref{fig11} \subref{fig11:b}, the empirical CDFs of V2V sum rate under the real-valued feedback and binary feedback are very close to each other and shift slightly to the left with increasing $\lambda_c$, which implies the rate degradation of V2V links is quite small. Besides, the CDFs of V2V sum rate under both feedback schemes are very close to that under the optimal scheme, and slightly deviate from the optimal performance with the further increase of $\lambda_c$. Thus, we can see that the proposed C-Decision scheme can ensure negligible degradation of V2V links while adjusting the QoS of V2I links via choosing different values of $\lambda_c$.

\subsection{Robustness Evaluation}
\begin{figure}
\begin{center}
\includegraphics[width=0.45\textwidth]{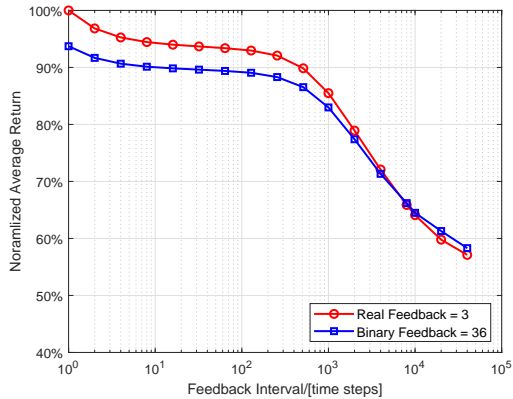}
\caption{Normalized average return with different feedback intervals.}
\label{Fig8}
\end{center}
\end{figure}

Fig. \ref{Fig8} shows the impacts of different feedback intervals on the performance of both real-valued feedback and binary feedback, where the feedback interval is measured in the number of testing steps. To investigate the impact of very large feedback intervals on the performance, we set the number of testing steps $T$ as $50,000$ and the number of testing episodes $L_{test}$ as $200$. The normalized average return under both feedback schemes decreases quite slowly with the increasing feedback interval at the beginning, which shows that the proposed scheme is immune to the feedback interval variations and then drops quickly with the very large feedback interval. Please note where the average return is normalized by the average return under the scheme with $N_k = 3$ since we set $T=50,000$ and it is very high computational demanding to find the return under the optimal scheme. 

\begin{figure} \centering
\subfigure[The ARP performance under the noisy input] {\label{Fig9:a}
\includegraphics[width=0.45\textwidth]{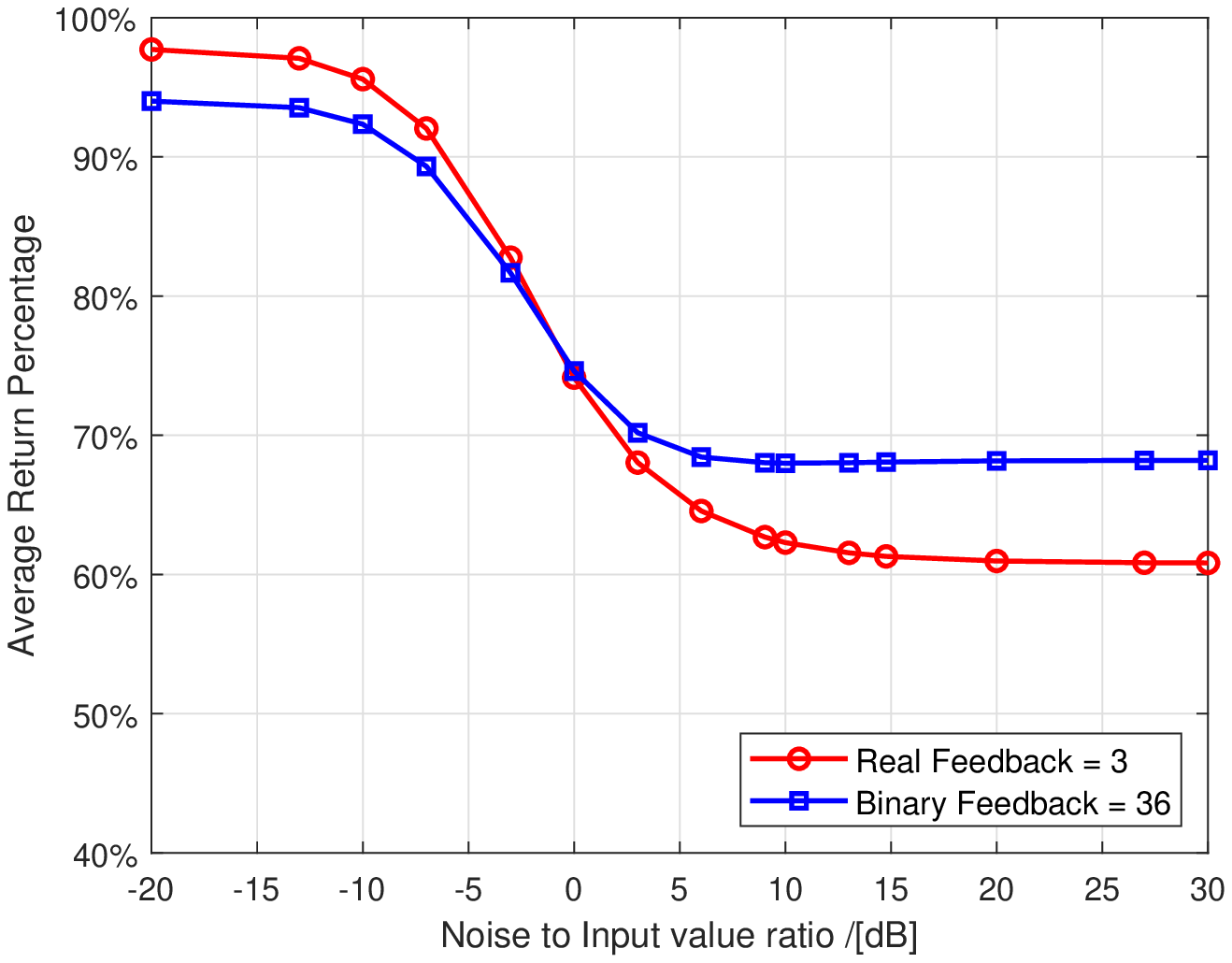}
}
\subfigure[The ARP performance under the noisy feedback] {\label{Fig9:b}
\includegraphics[width=0.45\textwidth]{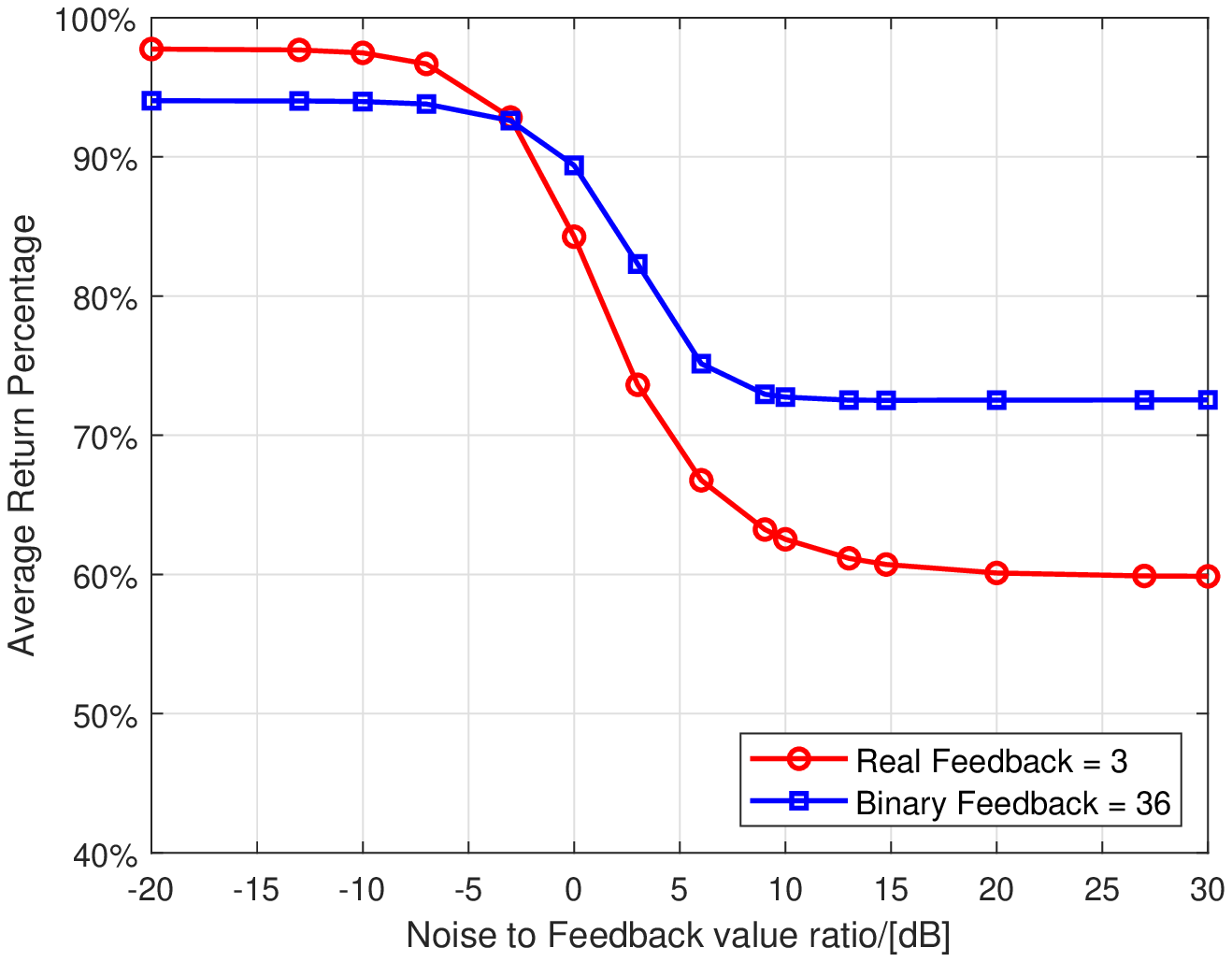}
}
\caption{Impact of noise on the ARP performance.}
\label{Fig9}
\end{figure}

Fig. \ref{Fig9} evaluates the impacts of different noise sources on the ARP performance. Specifically, Fig. \ref{Fig9} \subref{Fig9:a} illustrates the impacts of noisy input on the performance of both real-valued feedback and binary feedback. Here, the x-axis means the ratio of the strength of Gaussian white noise with respect to the each observation (such as channel gain value) for V2V links. In Fig. \ref{Fig9} \subref{Fig9:a}, the ARP under both feedback schemes decreases very slowly at the beginning and then drops very quickly, and finally keeps nearly unchanged with the very large input noise, which shows the robustness of the proposed scheme. In addition, the proposed scheme can also gain nearly $60\%$ of the optimal performance under both real-valued feedback and binary feedback even at the very large input noise, which is still better than the random action scheme shown in Fig. \ref{Fig4} \subref{Fig4:a}. Based on this observation, we remark the proposed scheme can learn the intrinsic structure of the resource allocation in the V2X scenario.



Besides, Fig. \ref{Fig9} \subref{Fig9:b} displays the impacts of noisy feedback on the performance of both feedback schemes. Here, noisy feedback refers to the situation where noise occurs when each V2V link sends its learned feedback to the BS. Similarly, the x-axis means the ratio of the strength of the Gaussian white noise with respect to each feedback. In Fig. \ref{Fig9} \subref{Fig9:b}, the ARP of both feedback schemes keeps nearly unchanged with the increasing feedback noise, which demonstrates the robustness of the proposed scheme, and then decreases more quickly under the real-valued feedback compared with that under the binary feedback with the further increasing feedback noise. This is because there are only $3$ real-valued feedback under the real-valued feedback scheme while there exist $36$ feedback bits under the binary feedback scheme. Finally, the ARP of both feedback schemes becomes nearly constant with the very large feedback noise. Similarly, the binary feedback scheme is more robust to the feedback noise compared with the real-valued feedback scheme. 


\subsection{Performance Evaluation for the D-Decision Scheme}

\begin{figure} \centering
\subfigure[Training loss of the $1$st V2V] { \label{Fig13:a}
\includegraphics[width=0.45\textwidth]{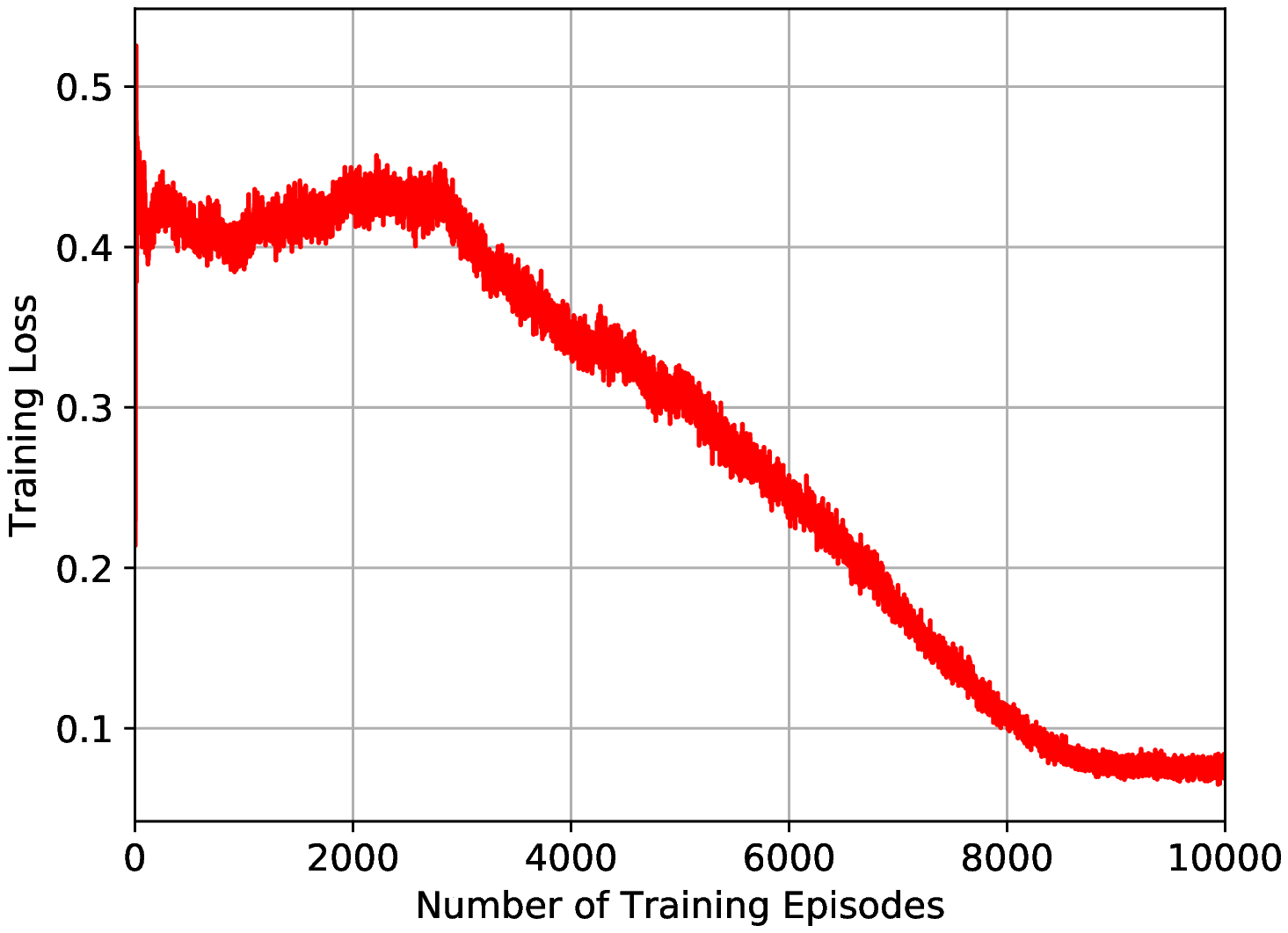}
}
\subfigure[Average return per episode] { \label{Fig13:b}
\includegraphics[width=0.45\textwidth]{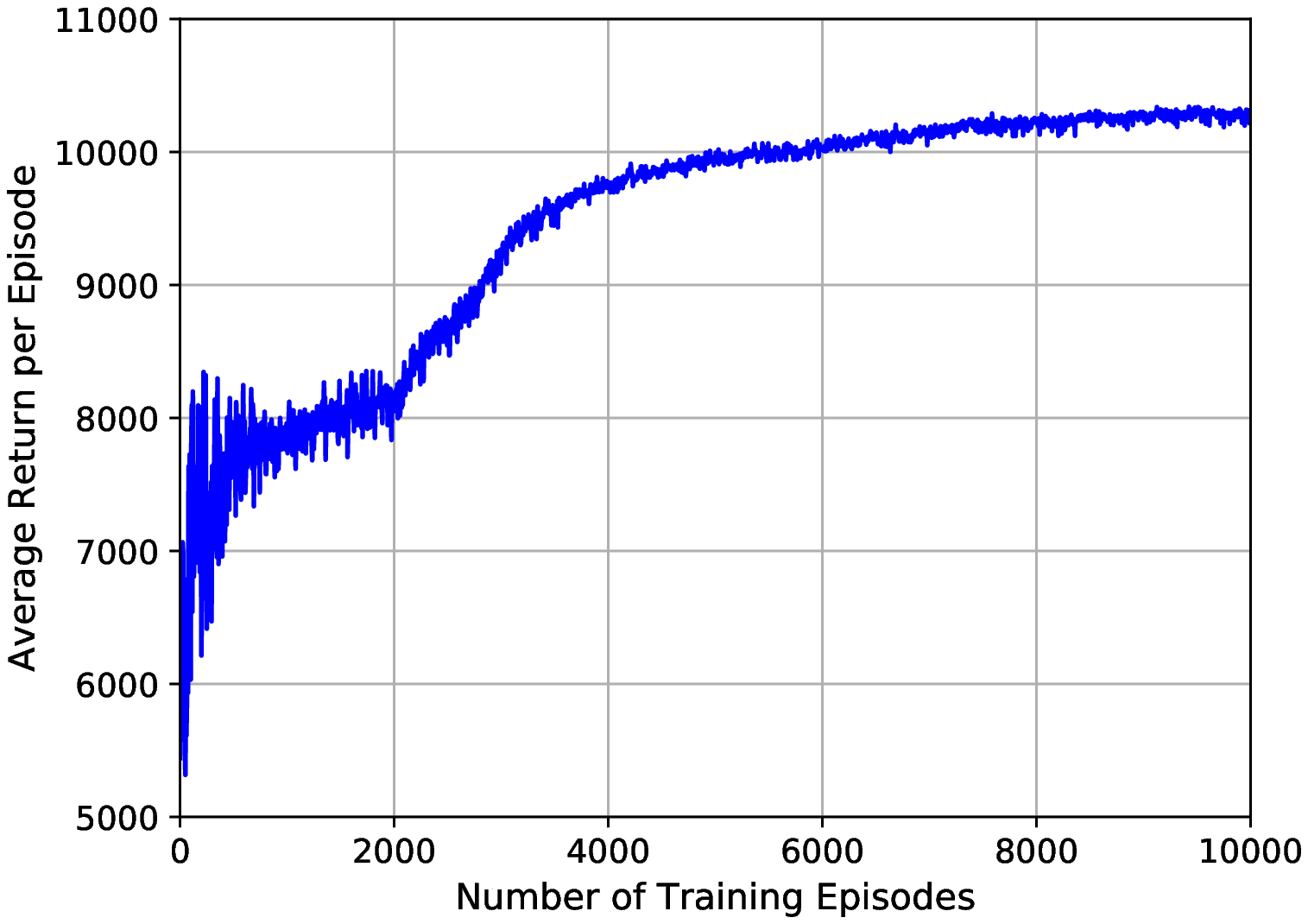}
}
\caption{Training performance evaluation for the D-Decision scheme.}
\label{Fig13}
\end{figure}

Fig. \ref{Fig13} evaluates the training process of the D-Decision scheme. Here, we choose $\mathcal{D} = 512$, $N_k = 3$ and $N_g^r = 16$, respectively. In particular, the training loss for the $1$st V2V in Fig. \ref{Fig13} \subref{Fig13:a} first decreases very slowly with some jitters with an increasing $L_{train}$, and then drops almost linearly, and finally becomes nearly unchanged with the further increase of $L_{train}$.  
The average return per episode under the D-Decision scheme in Fig. \ref{Fig13} \subref{Fig13:b} first increases quickly with the increase of $L_{train}$, and then increases slowly, and finally gradually converges despite some fluctuations, which shows the stability of the training process. Besides, we observe that $L_{train} = 10,000$ under the D-Decision scheme is much bigger than $L_{train} = 2,000$ under the C-Decision scheme, which indicates that the D-Decision scheme converges more slowly than the C-Decision scheme. To train the whole neural network well, we set $L_{train} = 10,000$ under the D-Decision scheme. Besides, the exploration rate $\epsilon$ is linearly annealed from $1$ to $0.01$ over the beginning $8,000$ episodes and then keeps constant. 

\begin{figure} \centering
\subfigure[ARP with real-valued aggregated global information] {\label{Fig12:a}
\includegraphics[width=0.45\textwidth]{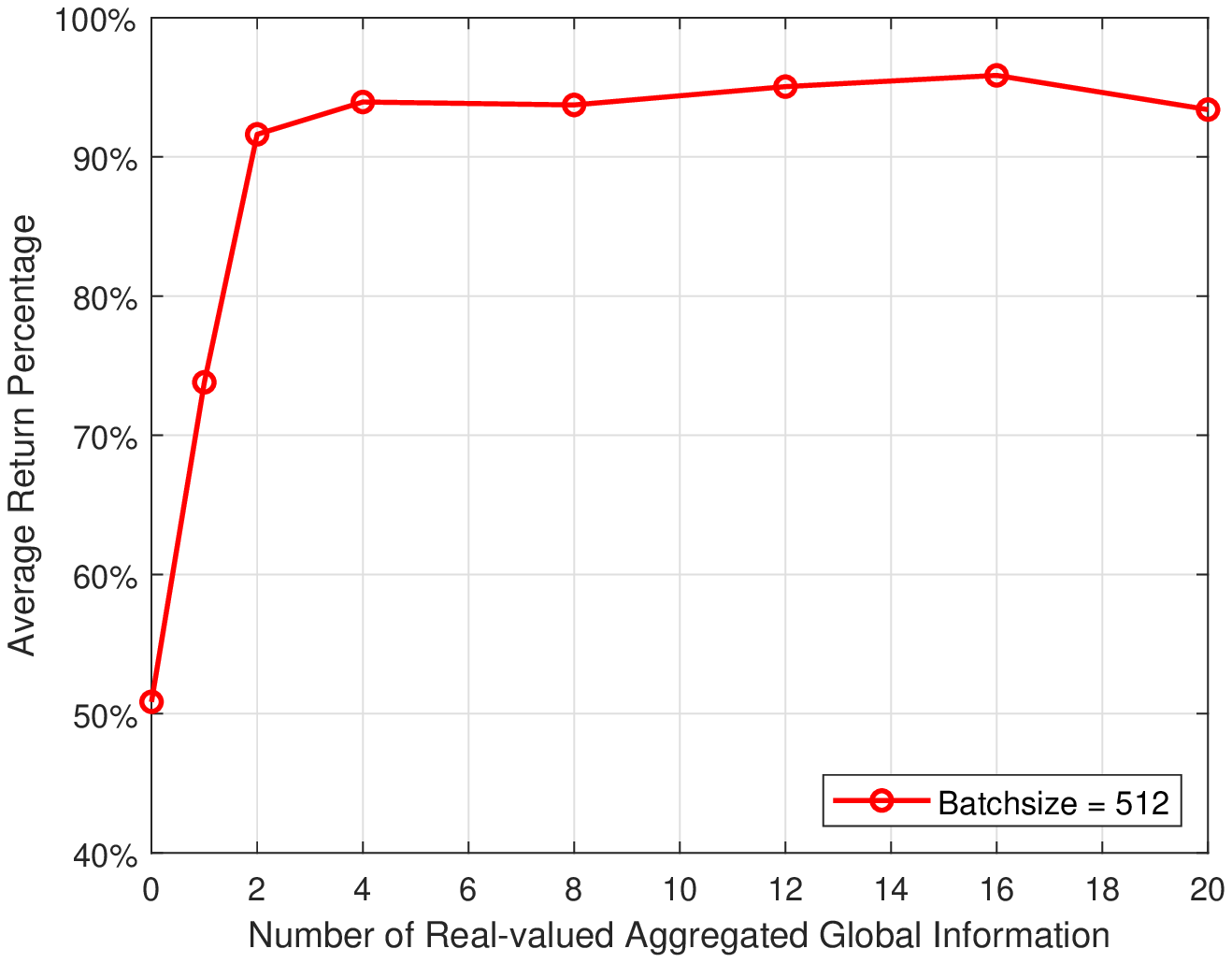}
} 
\subfigure[ARP with binary aggregated global information] {\label{Fig12:b}
\includegraphics[width=0.45\textwidth]{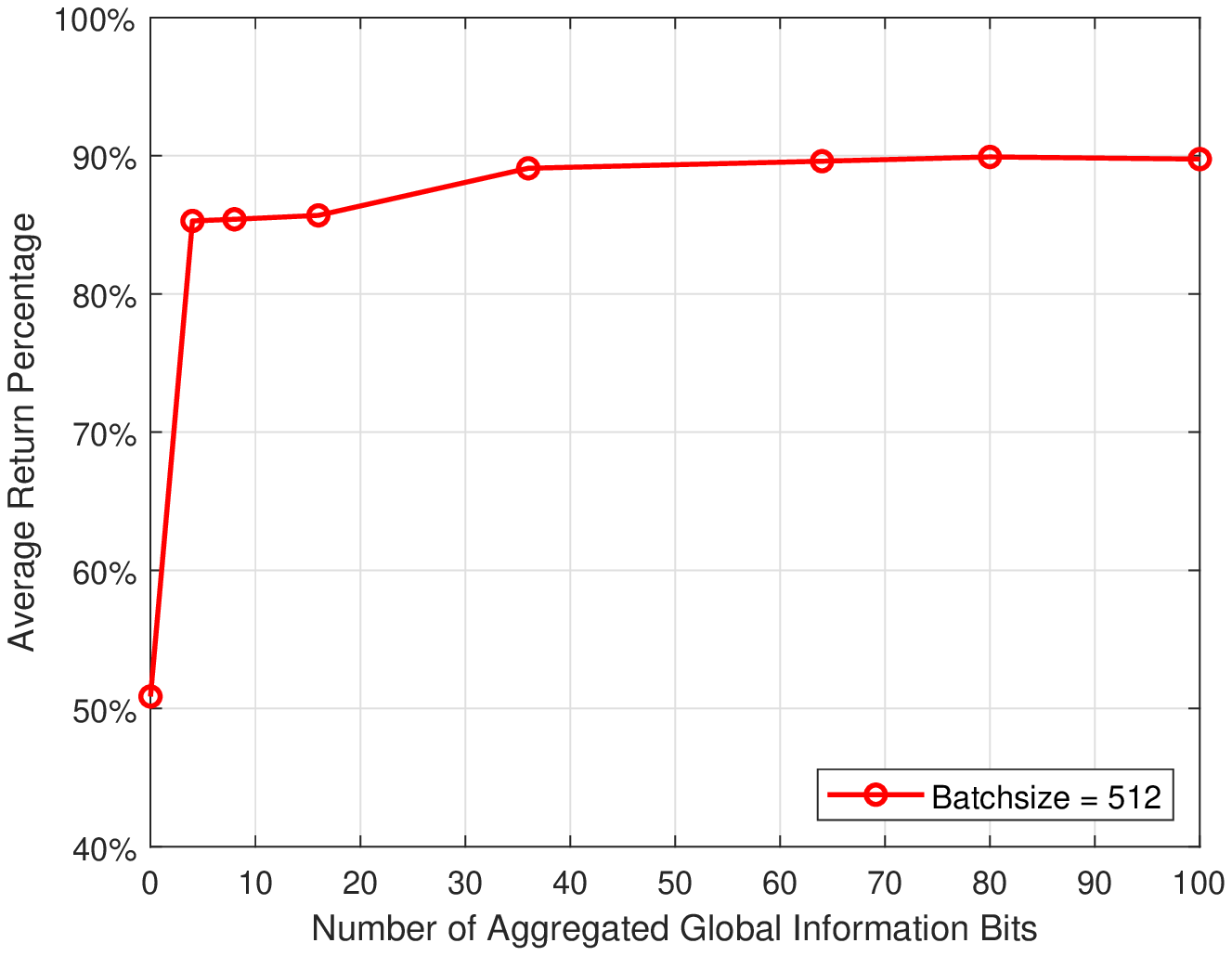}
}
\caption{The ARP performance with two kinds of aggregated global information.}
\label{Fig12}
\end{figure}

Then, the testing performance of the D-Decision scheme with the increasing number of AGI values is shown in Fig. \ref{Fig12}. In particular, Fig. \ref{Fig12} \subref{Fig12:a} illustrates the ARP performance with the increasing number of real-valued AGI $N_g^r$. Here, we set the number of real-valued feedback which each V2V transmits to the BS as $3$ as indicated by Fig. \ref{Fig4} \subref{Fig4:b}. The APR first increases with increasing $N_g^r$, and then keeps nearly unchanged with the further increase of $N_g^r$. Especially, the ARP nearly achieves its maximal value $96\%$ when $N_g^r = 16$. In other words, the BS only needs $16$ real-valued AGI to represent the real-valued feedback of all V2V links to achieve $96\%$ of the optimal performance. Furthermore, even when $N_g^r = 2$, the ARP can still reach $90\%$, which is suitable for the bandwidth-constrained broadcast channel of the BS. Compared with the C-Decision scheme, the D-Decision scheme only incurs $2\%$ ARP degradation. However, it can achieve the fully distributed decision making and spectrum sharing, which is very appealing in the V2X scenario. In addition, the computational complexity for decision making under the D-Decision scheme is greatly reduced compared with that under the C-Decision scheme, which can further facilitate the fully distributed spectrum sharing in the V2X scenario. 

Besides, the testing performance of the D-Decision scheme with the binary AGI is evaluated in Fig. \ref{Fig12} \subref{Fig12:b}. Here, we choose the number of feedback bits as $36$ for each V2V link and the number of real-valued AGI $N_g^r = 16$. In Fig. \ref{Fig12} \subref{Fig12:b}, the ARP first increases with the increasing number of AGI bits $N_g^b$, and then becomes nearly unchanged with the further increase of $N_g^b$. In particular, the APR reaches $90\%$ when $N_g^b = 80$. Meanwhile, the APR is very close to $90\%$ even when $N_g^b = 36$. Similarly, compared with the C-Decision scheme with binary feedback, the D-Decision scheme with the binary feedback only incurs $4\%$ ARP degradation, which, however, can be implemented in a fully distributed manner.

\section{Conclusion}
In this paper, we proposed a novel C-Decision architecture to allow distributed V2V links to share spectrum efficiently with the aid of the BS in V2X scenario and also devised an approach to binarize the continuous feedback. To further facilitate distributed decision making, we have developed a D-Decision scheme for each V2V link to make its own decision locally and also designed the binary procedure for this scheme. 
Simulation results demonstrated that the number of real-valued feedback can be quite small to achieve near-optimal performance. Meanwhile, the D-Decision scheme can also gain near-optimal performance and enable a fully distributed decision making, which is more appealing to the V2X networks.
Besides, the quantization of the feedback or AGI incurs small performance loss with an acceptable number of bits under both schemes. 
Our proposed scheme is quite immune to the variation of feedback interval, input noise, and feedback noise respectively, which validates the robustness of the proposed scheme. In the future, we will investigate joint power control and spectrum sharing issue in this scenario.



%

\bibliography{VtoVRef}

%
%
%

\end{document}